\documentclass[journal]{IEEEtran}

\usepackage{amsfonts, amsmath, amsthm}
\usepackage{graphicx, xspace, epsfig, syntonly, dsfont}
\usepackage{url, cite, bm, bbm}
\usepackage{algorithmicx}
\usepackage{balance}
\usepackage[caption=false]{subfig}
\usepackage{stfloats}
\usepackage{textcomp}
\usepackage{diagbox}
\usepackage{amsmath}

\usepackage{multirow}
\usepackage{graphicx}
\usepackage{amsmath}

\long\def\symbolfootnote[#1]#2{\begingroup
\def\thefootnote{\fnsymbol{footnote}}
\footnote[#1]{#2}\endgroup}

\def\BibTeX{{\rm B\kern-.05em{\sc i\kern-.025em b}\kern-.08em
    T\kern-.1667em\lower.7ex\hbox{E}\kern-.125emX}}

\begin{document}

\title{Physics-guided Residual Learning for \\ Probabilistic Power Flow Analysis}

\author{
\IEEEauthorblockN{Kejun Chen and Yu Zhang,~\IEEEmembership{Member,~IEEE}}

\thanks{K. Chen and Y. Zhang are with the Department of Electrical and Computer Engineering at the University of California, Santa Cruz. Emails: \texttt{\{kchen158,yzhan419\}@ucsc.edu}
This work was supported in part by the Faculty Research Grant of UC Santa Cruz and the Hellman Fellowship (\emph{Corresponding author: Yu Zhang}).}
}
\maketitle

\begin{abstract}
Probabilistic power flow (PPF) analysis is critical to power system operation and planning. PPF aims at obtaining probabilistic descriptions of the state of the system with stochastic power injections (e.g., renewable power generation and load demands). Given power injection samples, numerical methods repeatedly run classic power flow (PF) solvers to find the voltage phasors. However, the computational burden is heavy due to many PF simulations. Recently, many data-driven based PF solvers have been proposed due to the availability of sufficient measurements. This paper proposes a novel neural network (NN) framework which can accurately approximate the non-linear AC-PF equations. The trained NN works as a rapid PF solver, significantly reducing the heavy computational burden in classic PPF analysis. Inspired by residual learning, we develop a fully connected linear layer between the input and output in the multilayer perceptron (MLP). To improve the NN training convergence, we propose three schemes to initialize the NN weights of the shortcut connection layer based on the physical characteristics of AC-PF equations. Specifically, two model-based methods require the knowledge of system topology and line parameters, while the purely data-driven method can work without power grid parameters. Numerical tests on five benchmark systems show that our proposed approaches achieve higher accuracy in estimating voltage phasors than existing methods. In addition, three meticulously designed initialization schemes help the NN training process converge faster, which is appealing under limited training time.
\end{abstract}


\section{Introduction}\label{sect:intro}
Renewable power generation technology has been developed rapidly due to its great advantage in economic savings and environmental friendliness \cite{Bazilian2013}. However, compared with conventional generation, renewable generation (e.g., solar and wind power) is highly dependent on weather conditions, such as ambient temperature, solar irradiance, relative humidity, wind speed, etc \cite{FU2017}. Hence, renewable generation brings lots of uncertainties to power system operation, e.g., significant fluctuations of voltage phasors and branch flows. Probabilistic power flow (PPF) analysis is essential for characterizing power system uncertainties under various random variations \cite{Borkowska1974_ppf}. PPF focuses on obtaining the probabilistic distributions of voltage phasors and branch flows (output variables) under the uncertainties of power injections (input variables). Existing schemes for solving PPF problems can be divided into analytical, approximate, and numerical methods. 

Analytical methods use the first-order Taylor expansion of AC power flow (AC-PF) equations around the operating point. The uncertainties of voltage phasors (including voltage magnitudes and voltage angles) are represented as linear combinations of the variations of power injections. For example, cumulant methods use arithmetic operations to calculate moments of voltage phasors and obtain their probability density functions (PDFs) with various series expansions \cite{Miao2012}. Their computational efficiency is high for large-scale power systems. However, the linearization approximation suffers significant errors when power injections deviate far from the operating point \cite{Deng2017}. In addition, these series expansions (e.g., Gram–Charlier, Edgeworth, and Cornish–Fisher expansions) cannot always guarantee convergence \cite{Miao2012}.  

Point estimation methods are the most commonly used approximate methods, which conduct deterministic PF analysis over a limited number of selected sample points \cite{Juan2007}. Given the first few statistical moments of input variables, they calculate the moments of output variables and then apply series expansions to obtain their PDFs. However, calculating exact higher‐order moments of many input variables is challenging, leading to decreased estimation accuracy in medium- and large-scale bus systems \cite{Che2019}. In addition, to cope with the potential divergence issue brought about by the series expansions, the stochastic response surface method has been applied to PPF analysis. It can describe the system responses based on polynomial chaos expansion and obtain the PDFs using kernel density estimation, e.g., \cite{Fu2020} and \cite{Fu2022}. 

Numerical methods, whose canonical example is Monte Carlo simulation (MCS), have been widely applied in PPF analysis \cite{Constante2019}. MCS generates samples from stochastic power injection distributions and calculates their corresponding voltage phasors using the Newton–Raphson (NR) algorithm. In addition, some improved MCS-like approaches achieve better computation efficiency by reducing the sampling number, such as Latin supercube sampling \cite{Mahdi2013} and Quasi-Monte Carlo (QMC) \cite{Xu2017}. However, these improved MCS-like methods still need to call the NR solver repeatedly over many samples to guarantee accurate estimation of the PDFs of voltage phasors. Therefore, the cumulative computational time of these samples is long. 

Speeding up the computation of PF analysis of each sample without sacrificing accuracy is another promising direction to reduce the total computational time. Recently, the widespread use of massive phasor measurement units (PMUs) and supervisory control and data acquisition (SCADA) systems can collect sufficient measurement data. Thus, data-driven PF solvers have gained increasing attention recently. They learn the mapping from power injections to voltage phasors based on historical input-output data pairs \cite{Yu2017}. For example, \cite{Liu2019} and \cite{Liu2020} use linear regression to approximate the decoupled linear PF functions. Their proposed approaches do not rely on power grid parameters and perform better than model-based decoupled linear PF models. However, these linear models suffer from accuracy limitations because they cannot extract non-linear features of the PF functions. In addition, \cite{Xu2020} and \cite{Pareek2022} use Gaussian process regression, which can only obtain the PDF of a single target quantity in one shot. This property prevents its application in medium- and large-scale power systems if the PDFs of all buses' voltage phasors are required. 

Exploiting the impressive capability of neural networks (NNs) in function approximation, we focus on employing the NN as a rapid PF solver. The motivations are primarily two-fold. First, the universal approximation theorem states that NNs can approximate any arbitrary complex functions \cite{HORNIK1989}. Thus, they are expected to approximate the AC-PF equations accurately. Secondly, the time-consuming training process is implemented offline. Once the training is completed, the trained NN is ready to participate in PF analysis. Specifically, they take power injections as the input and output the corresponding voltage phasors. The forward propagation is fast; therefore, the total prediction time of lots of input power injection samples is negligible. The PDFs of voltage phasors based on the output samples can be further obtained.

NN-based approaches have recently become popular in PF and PPF analysis. For example, \cite{Yang2020} and \cite{Xiang2020} employ multilayer perceptrons (MLPs) to approximate the AC-PF equations and achieve much less computational time than MCS. The loss function consists of the mean square error of voltage magnitudes/angles and active/reactive branch flows. However, the capability of NN may be insufficient to minimize these four errors simultaneously \cite{Kejun2022}. Besides, the NN training time will significantly increase due to the branch flow calculations involvement. In addition, \cite{Hu2021} employs MLPs while incorporating the system topology to solve the PF problem. Under physical guidance, their proposed topology-pruned bilinear neural network (TPBNN) method performs better than the MLP. However, the model outputs are the voltage phasors' real and imaginary parts. After transforming to voltage magnitudes and angles, the calculation errors can get magnified. In addition, electrical grids can be abstracted as sparse-connected graphs composed of nodal buses and power branches. Graph convolutional neural networks can exploit topological information and aggregate locality information. \cite{Bolz2019} and \cite{Wang2020} have applied them to PF analysis. However, the formulation of AC-PF equations shows that the power injections of each bus will affect not only its neighboring buses but all other buses in the electrical grid. Therefore, the benefits of graph convolutional layers over fully connected layers may be obscure \cite {Falconer2022}.

In addition, residual neural networks (ResNets) have been widely used in image recognition. ResNets consist of many stacked residual building blocks, which are achieved by introducing skip-layer connections among layers in MLPs. A central choice is attaching a direct path between the input and output layers in one residual block, the so-called identity shortcut connection. These shortcut connections can help deal with vanishing gradients and accuracy saturation in deep NNs \cite{Veit2016}. The motivation is that learning residuals regarding identity mapping should be easier than learning the desired mapping directly \cite{He2016}. 

In this paper, we propose a novel physics-guided NN framework to approximate the inverse AC-PF equations. We further employ it as the rapid PF solver to reduce the total computational time of many samples in PPF analysis. The PDF can be obtained using the non-parametric kernel density estimation method based on data samples. The main contributions of this paper are summarized below: 
\begin{itemize}
    \item Given sufficient historical operational input-output data pairs, we train the NN to learn the mapping from power injections to voltage phasors. The trained NN will serve as a rapid data-driven PF solver in PPF analysis.
    
    \item Motivated by residual learning, we introduce a fully connected linear layer between the input and output in the MLP structure. Therefore, the original MLP essentially learns the non-linear corrections to the linear mapping from the power injections to voltage phasors.
    
    \item Leveraging the linearized formulations of AC-PF equations, we design three initialization methods for the weights of the shortcut connection layer. Furthermore, the proposed data-driven initialization scheme does not require the knowledge of system topology and line parameters. 
    
\end{itemize}

The remaining part of this paper is organized as follows. Section \ref{sec:pf} describes the problem formulation. Section \ref{sec:Model based residual network} details the proposed network and three novel initialization methods. Section \ref{sec:Numerical Results} shows the simulation results tested on five benchmark systems. Finally, Section \ref{sec:conclusion} presents the concluding remarks.

\textit{Notation}: Upper (lower) boldface letters are used for matrices (column vectors). Sets are denoted by calligraphic letters. $(\cdot)^{\top}$ is vector/matrix transpose; $\|\cdot\|_2$ denotes vector $\ell_2$-norm; $(\cdot)^{-1}$ and $(\cdot)^{\dagger}$ denote inverse and pseudo-inverse, respectively.

\section{Problem Formulation} \label{sec:pf}
Deterministic PF analysis is the cornerstone of PPF analysis. This section first introduces the system description and the problem formulation. Then, we show the connection between PF analysis and PPF analysis. 

\subsection{System Description} PF analysis aims to analyze the steady-state operating points of an electrical grid. The operational data includes power generation, load demands, voltage phasors, and branch flows. There are three types of buses: PQ buses, PV buses, and one slack bus. A PQ bus (a.k.a. load bus) has no generator attached, where its active and reactive power injections are fixed. A PV bus (generator bus) has generators connected, and its active power injection and voltage magnitude are known. Lastly, the slack bus has the given voltage angle and voltage magnitude. Therefore, the unknown variables include the voltage angles and voltage magnitudes of PQ buses and the voltage angles of PV buses. 

Consider a power grid with $N$ buses, where there are one slack bus, $N_g$ PV buses (denoted by set $\mathcal{N}_{g}$), and $(N-N_g-1)$ PQ buses (denoted by set $\mathcal{N}_{l}$). The number of unknown variables is $2\times (N-N_g-1) + N_g$, which is the same as the number of power balance equations.

\subsection{AC Power Flow Equations} 
Let $P_i$ and $Q_i$ represent the active and reactive power injections at bus $i$ while $V_i$ and $\theta_i$ are the voltage magnitude and angle.
The AC-PF equations are the forward mappings from voltage phasors to power injections, which are given as:
\begin{subequations} 
\label{eq:ACPF}
  \begin{align}
    P_i &= \sum_{j=1}^N V_i V_j(G_{ij}\cos{\theta_{ij}}+B_{ij}\sin\theta_{ij}),\, \forall i\in \mathcal{N}_{g} \cup \mathcal{N}_{l}\, , \label{ACPF1} \\ 
    Q_i &= \sum_{j=1}^N V_i V_j(G_{ij}\sin{\theta_{ij}}-B_{ij}\cos\theta_{ij}),\, \forall i\in \mathcal{N}_{l}\, , \label{ACPF2}
   \end{align}  
\end{subequations}
where $\theta_{ij} := \theta_i - \theta_j$ is the voltage angle difference between bus $i$ and bus $j$. $G_{ij}$ and $B_{ij}$ are the real and imaginary parts of the $(i,j)$-th element of the nodal admittance matrix $\mathbf{Y} \in \mathbb{C}^{N \times N}$. Moreover, the active and reactive power branch flows between two connected buses $i$ and $j$ are:
\begin{subequations}
\label{eq:branchflow}
\begin{align}
    P_{ij} &= V_iV_j(G_{ij}\cos{\theta_{ij}}+B_{ij}\sin\theta_{ij}) - G_{ij}V_i^2\, , \label{BF1} \\
    Q_{ij} &= V_iV_j(G_{ij}\sin{\theta_{ij}}-B_{ij}\cos\theta_{ij}) + B_{ij}V_i^2 - \frac{b_{ij}^c}{2}V_i^2\, ,
    \label{BF2}
\end{align}
\end{subequations}
where $b_{ij}^c$ is the total line-charging susceptance.

Clearly, the power injections and branch flows are determined by all the voltage phasors, which are defined as the state of the system:
\begin{equation*}
   \mathbf{z}:= [\theta_{s}, \bm{\theta}_{g}, \bm{\theta}_{l},
V_s, \mathbf{V}_g, \mathbf{V}_l]^{\top}\, , 
\end{equation*}
where subscripts $(\cdot)_{s}$, $(\cdot)_{g}$ and $(\cdot)_{l}$ denote the quantities corresponding to the slack bus, generator buses, and load buses, respectively. Let $\mathbf{P}_{g}$, $\mathbf{P}_{l}$, and $\mathbf{Q}_{l}$ denote the 
active power of generator buses and 
active/reactive power injections of load buses. Partitioning and reorganizing the admittance matrix $\mathbf{Y}$ in the same manner yielding
\begin{align}\label{admitance}
    \tilde{\mathbf{Y}} = \left[
  \begin{array}{ccc}
    \mathbf{Y}_{ss} & \mathbf{Y}_{sg} & \mathbf{Y}_{sl} \\
    \mathbf{Y}_{gs} & \mathbf{Y}_{gg} & \mathbf{Y}_{gl}  \\
    \mathbf{Y}_{ls} & \mathbf{Y}_{lg} & \mathbf{Y}_{ll} \\
  \end{array}
\right]\, ,
\end{align}
where $\mathbf{Y}_{gs}$ is formed from the rows of $\mathbf{Y}$ that correspond to generator buses and the column for the slack bus. Similarly, we can find all other blocks. Define 
\begin{align*}
\mathbf{x} = [\mathbf{P}_g; \mathbf{P}_l; \mathbf{Q}_l]^{\top},\\ 
\mathbf{y} = [\bm{\theta}_g; \bm{\theta}_l; \mathbf{V}_l]^{\top}.
\end{align*}
Hence, the inverse mappings of the AC-PF equations can be compactly expressed as:
\begin{align}\label{eq:inversePFmapping}
    \mathbf{y} = \mathbf{f}(\mathbf{x})\, .
\end{align}

\subsection{PPF Analysis}
The PPF analysis is closely related to the aforementioned AC-PF problem. Consider an electricity grid with stochastic renewable power generation and load demands. PPF studies aim to characterize the uncertainties of voltage phasors induced by the fluctuations of power injections. As a new effort in PPF studies, NNs are employed to learn the end-to-end mapping \eqref{eq:inversePFmapping} from historical data pairs $(\mathbf{x}, \mathbf{y})$. NNs take in power injections $\mathbf{x}$ and predict the corresponding voltage phasors $\mathbf{y}$. The PDFs of voltage phasors can be further inferred from the output samples. By shifting the time-consuming training process offline, NNs can rapidly predict the corresponding voltage phasors of new power injection samples in the testing stage. Even with lots of new input samples, NNs are still computational efficiently because the forward propagations during the test stage are often very fast.

\section{Physics-guided PPF Analysis} \label{sec:Model based residual network}
This section presents the details of our proposed NN structure designed for approximating the inverse AC-PF mappings \eqref{eq:inversePFmapping}. Furthermore, we develop three novel initialization methods for the shortcut connection layer in different scenarios. Two are model-based initialization schemes, which require the knowledge of system topology and line parameters, while the other data-driven initialization scheme does not require the power grid's parameters. 

\subsection{Residual Learning} 
\begin{figure}[t]
    \centering
    \subfloat[\label{fig:residual_block}]{%
    \includegraphics[scale=0.06]{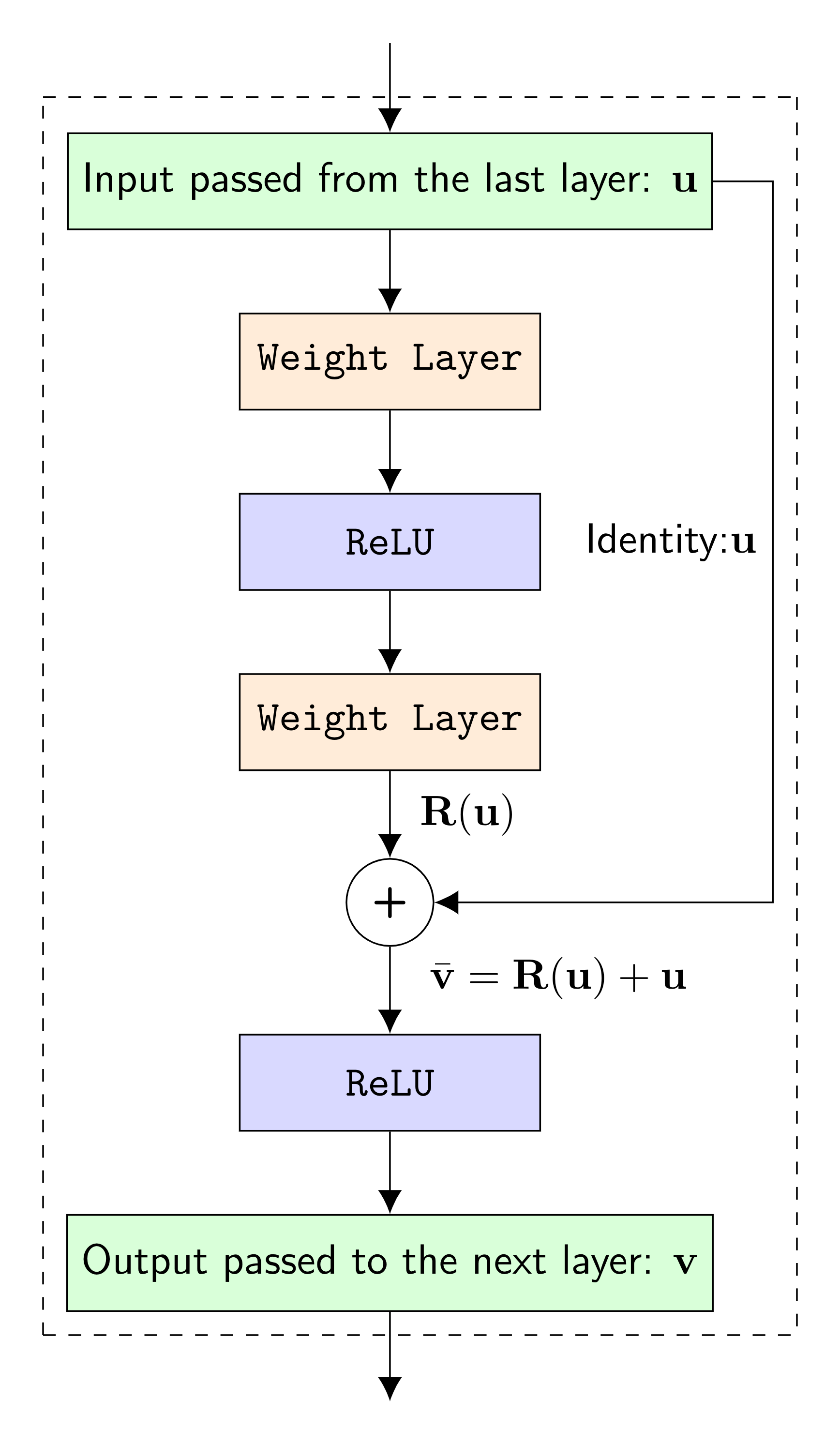}}
    \hfill
    \subfloat[\label{fig:residual_block1}]{%
    \includegraphics[scale=0.047]{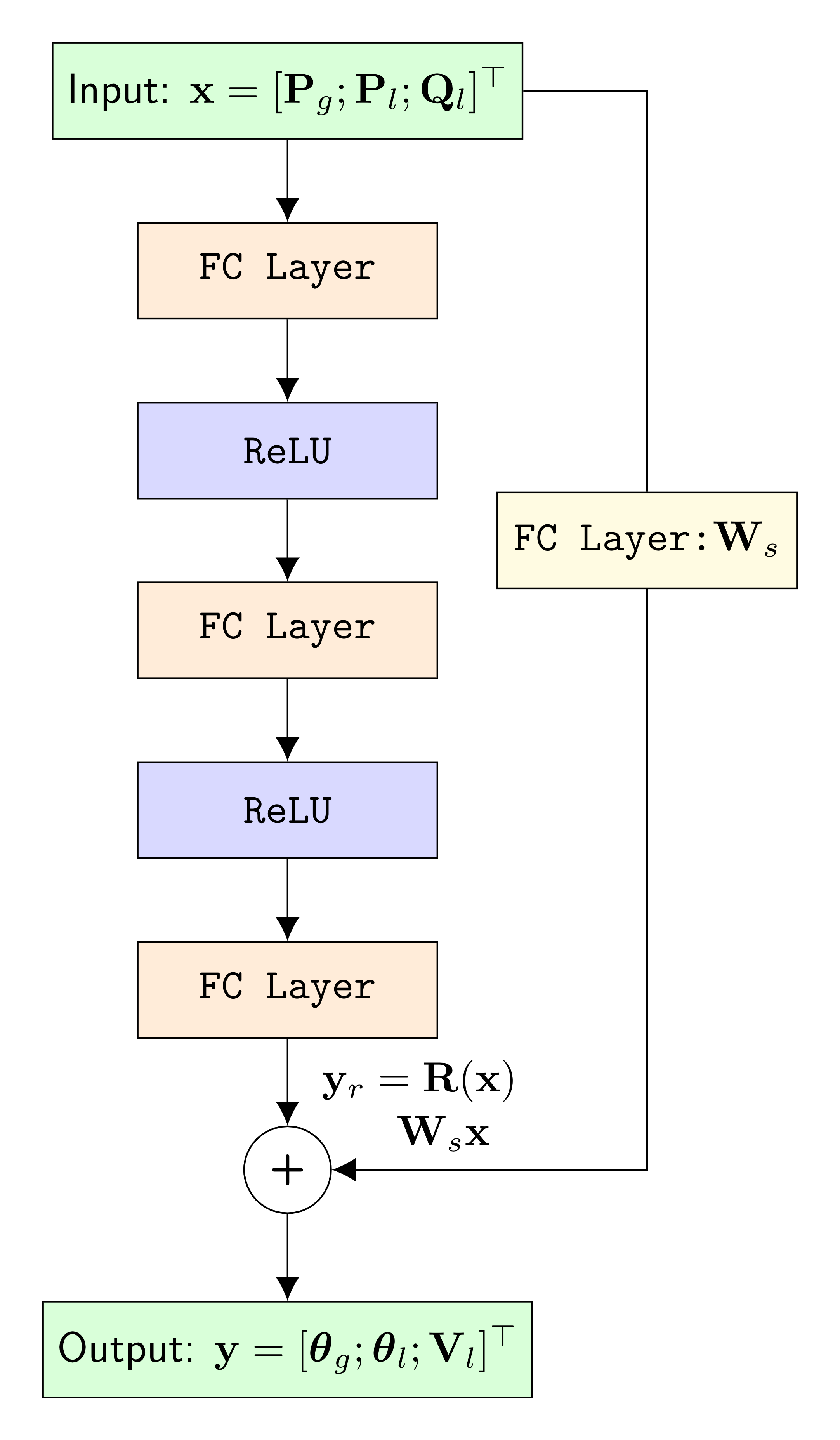}}
    \caption{(a) One residual building block that skips two weight layers. (b) The proposed architecture tailored to the PF analysis.}
  \label{fig1} 
\end{figure}

Unlike the building blocks in the MLP, residual blocks introduce identity mapping as a shortcut connection that skips one or more layers \cite{He2016}. Fig.~\ref{fig:residual_block} illustrates the structure of one residual building block. In Fig.~\ref{fig:residual_block}, let $\mathbf{G}(\cdot)$ denote the mapping from $\mathbf{u}$ to $\mathbf{\bar{v}}$. The motivation for formulating the residual building block is as follows. MLP consists of a few stacked layers and shows universal function approximation capabilities. Thus, it is reasonable to assume that the MLP can approximate the residual function $\mathbf{R}(\cdot)$. Then, if those stacked layers are only used to approximate $\mathbf{R}(\cdot)$, the original function $\mathbf{G}(\mathbf{u}):= \mathbf{R}(\mathbf{u}) + \mathbf{u}$ can be obtained by introducing an extra identity shortcut connection. It has been shown that ResNets are also universal function approximators \cite{Lin2018}.

\subsection{Designed framework for PF Analysis} 
However, the benefits of ResNet over MLP in PF analysis are not obvious. The reason is as follows. The NNs applied in the image recognition domain are usually composed of dozens or hundreds of layers. Therefore, deep ResNets will have apparent benefits over deep MLPs because they do not have degradation issues. However, the advantages are not evident because very deep NNs may not be necessary for approximating the inverse AC-PF equations.  

In addition, \cite{He2016} claimed that if the desired function is close to the identity function, it will be easier for stacked layers to learn the perturbations regarding the identity instead of approximating that desired function directly. Inspired by their work, our proposition is that compared with MLPs that directly approximate the complicated inverse AC-PF equations, pushing the residual correction regarding the linear relationship to zero may be more accessible. This motivates us to design a novel architecture with some initialization methods tailored for PF analysis.

In practical power systems, the voltage magnitudes of buses are approximately 1 per unit, and voltage angle differences are small values \cite{Fatemi2015}. Many existing works use these two properties to linearize the AC-PF equations, and numerical results have shown that these linear models perform well. Therefore, we make novel modifications to the MLP structure to better use this linear property. As shown in Fig.~\ref{fig:residual_block1}, we replace the identity mapping with a fully connected linear layer. Compared with the MLP that tries to approximate the mapping from $\mathbf{x}$ to $\mathbf{y}$ directly, our designed framework uses a set of stacked layers to only approximate the latent residual functions $\mathbf{R}(\cdot)$. The stacked layers are composed of fully connected linear layers and rectified linear unit (ReLU) activation function \cite{glorot2011}. Let $\mathbf{W}_s$ and $\mathbf{b}_s$ denote the weight matrix and bias of the shortcut connection linear layer, respectively. $\mathbf{W}_i$ and $\mathbf{b}_i$ are the weight matrix and bias of the $i$-th fully connected linear layer. $\boldsymbol{\sigma}$ is the activation function. The residual output $\mathbf{y}_r$ and the final output $\mathbf{y}$ can be written as:
\begin{align}\label{residual variable}
    \mathbf{y}_r &= \mathbf{W}_3(\boldsymbol{\sigma} (\mathbf{W}_2( \boldsymbol{\sigma}(\mathbf{W}_1\mathbf{x}+\mathbf{b}_1))+\mathbf{b}_2)) + \mathbf{b}_3\, ,  \\
    \mathbf{y} &= \mathbf{y}_r + (\mathbf{W}_s\mathbf{x}+ \mathbf{b}_s)\, .
\end{align}

\subsection{Physics-guided Initialization Methods} \label{Initialization Methods}
In NN training, learnable weights are randomly initialized which can be critical to the NN performance. We develop two physics-guided initialization methods for the weights of the shortcut connection layer. These two methods significantly accelerate the training process and improve the NN performance than random initialization. The goal is to get $\mathbf{y}_r$ close to zero in the initial stage of the training process by initializing $\mathbf{W}_s$ and $\mathbf{b}_s$ properly. If $\mathbf{W}_s$ and $\mathbf{b}_s$ are randomly initialized, the value of $\mathbf{y}_r = \mathbf{y} - (\mathbf{W}_s\mathbf{x}+\mathbf{b}_s)$ cannot be zero. Thus, to drive $\mathbf{y}_r$ to zero, $\mathbf{W}_s\mathbf{x} + \mathbf{b}_s$ should be close to the output $\mathbf{y}$. 
In other words,
the shortcut connection linear layer should serve as an excellent linear approximation from $\mathbf{x}$ to $\mathbf{y}$ after initializing its weights. Therefore, we modify two linear PF models and apply them to initialize $\mathbf{W}_s$ and $\mathbf{b}_s$ in the proposed framework. 

\subsubsection{Pre-initialization using linearized PF model}
A decoupled linear PF model is proposed in \cite{Yang2017}, and the linearized AC-PF equations can be expressed as:
\begin{subequations}
\label{eq:DLPF}
\begin{align}
    P_i &= \displaystyle\sum_{j=1}^N -B_{ij}^{\prime}\theta_j + \displaystyle\sum_{j=1}^NG_{ij}V_j ,\, i\in \mathcal{N}_{g} \cup \mathcal{N}_{l}, \label{DLPF1} \\
    Q_i &=  \displaystyle\sum_{j=1}^N -G_{ij}\theta_j +  \displaystyle\sum_{j=1}^N-B_{ij}V_j,\, i\in \mathcal{N}_{l}, \label{DLPF2}
\end{align} 
\end{subequations}
where $B_{ij}^{\prime}$ is the imaginary part of the $(i,j)$-th element of the nodal admittance matrix without shunt elements and line-charging susceptance. Note that the summation in equation \eqref{eq:DLPF} contains all the buses. Since the voltage phasor of the slack bus and the voltage magnitudes of PV buses are known, we separate them from the remaining unknown voltage phasors. After the separation, \eqref{eq:DLPF} can be compactly rewritten as
\begin{align} \label{LPF}
    \mathbf{x} = \mathbf{E}\mathbf{c} + \mathbf{F}\mathbf{y}\, ,
\end{align}
where
\begin{align*}
    \mathbf{E} &= \left[
  \begin{array}{ccc}
    -\mathbf{B}_{gs}^{\prime} & \mathbf{G}_{gs} & \mathbf{G}_{gg} \\
    -\mathbf{B}_{ls}^{\prime} & \mathbf{G}_{ls} & \mathbf{G}_{lg}  \\
    -\mathbf{G}_{ls} & -\mathbf{B}_{ls} & -\mathbf{B}_{lg}  \\
  \end{array}
\right] \in \mathbb{R}^{(2N-N_g-2)\times(N_g+2)}\, ,\\
\mathbf{F} &= \left[
  \begin{array}{ccc}
    -\mathbf{B}_{gg}^{\prime} & -\mathbf{B}_{gl}^{\prime} & \mathbf{G}_{gl} \\
    -\mathbf{B}_{lg}^{\prime} & -\mathbf{B}_{ll}^{\prime} & \mathbf{G}_{ll}  \\
    -\mathbf{G}_{lg}     & -\mathbf{G}_{ll} & -\mathbf{B}_{ll} \\
  \end{array}
\right] \in \mathbb{R}^{(2N-N_g-2)\times (2N-N_g-2)}, \\
\mathbf{c} &= 
[{\theta}_s, {V}_s,\mathbf{V}_g]^{\top} \in \mathbb{R}^{N_g+2}\, .
\end{align*}

Assuming that the network topology and line parameters are known, matrices $\mathbf{E}$ and $\mathbf{F}$ are fixed. Vector $\mathbf{c}$ is also fixed since the voltage phasor of the slack bus and voltage magnitudes of PV buses are known. According to \eqref{LPF}, the linear mapping from $\mathbf{x}$ to $\mathbf{y}$ becomes:
\begin{equation}
    \mathbf{y} = \mathbf{F}^{\dagger}\mathbf{x} - \mathbf{F}^{\dagger}\mathbf{E}\mathbf{c}\, ,
\end{equation}
where $\mathbf{F}^{\dagger}$ is the pseudo-inverse of $\mathbf{F}$. Note that if there are zero bus injections for the PV and PQ buses, $\mathbf{F}$ is not invertible. Thus, we can use $\mathbf{F}^{\dagger}$ and $- \mathbf{F}^{\dagger}\mathbf{E}\mathbf{c}$ to pre-initialize $\mathbf{W}_s$ and $\mathbf{b}_s$, respectively.

\subsubsection{Pre-initialization using Jacobian matrix model} AC-PF equations can be linearized based on the first-order Taylor expansion around the nominal operating point, denoted as $(\mathbf{x}_0 \, , \mathbf{y}_0)$. Expanding \eqref{eq:ACPF} around the operating point and ignoring the higher-order terms, the linearized AC-PF equations can be expressed as:
\begin{align}
    \mathbf{x} &= \mathbf{x}_0 + \mathbf{J}(\mathbf{y}-\mathbf{y}_0)\, ,\\
    \mathbf{y} &= \mathbf{J}^{-1}\mathbf{x} - \mathbf{J}^{-1}\mathbf{x}_0 + \mathbf{y}_0 \, ,
\end{align}
where the Jacobian matrix $\mathbf{J} \in \mathbb{R}^{(2N-N_g-2)\times (2N-N_g-2)}$ is evaluated at $\mathbf{y}_0$. We can use $\mathbf{J}^{-1}$ and $-\mathbf{J}^{-1}\mathbf{x}_0 + \mathbf{y}_0$ to pre-initialize $\mathbf{W}_s$ and $\mathbf{b}_s$.

The reason for using the Jacobian matrix for pre-initialization is straightforward. If the input power injections are slightly perturbed around $\mathbf{x}_0$, their corresponding output voltage phasors should be close to $\mathbf{y}_0$. In the first training iteration, the residual output represents the higher-order remainder of the Taylor series. Thus, the value should be closer to zero than that under the random initialization method. 

\subsection{Initialization based on Data-driven PF Model}

The line parameter profiles are only available from the grid planning files, which are likely outdated \cite{Hu2021}. If accurate information on line parameters is unavailable, the aforementioned physics-guided initialization methods are not applicable. Therefore, we propose a data-driven initialization scheme by leveraging ridge regression to deal with this situation. The motivation for using ridge regression is as follows. First, \cite{Junbo2016} shows the load demands and power generation data have similar rise and fall patterns in adjacent areas. The high correlations among the input variables lead to multicollinearity, weakening the performance of the simple linear regression model. Ridge regression is an effective method for analyzing data that suffer from multicollinearity \cite{schreiber2018}. It introduces the $\ell_2$ regularization, which shrinks the regression coefficients to reduce model variance and avoid overfitting. Secondly, the initial weight parameters of NNs are typically random numbers following certain distributions. Small random values are generally considered better options than larger ones in preventing gradient from exploding or vanishing \cite{Hanin2018}. Due to the penalty term, ridge regression tends to give smaller random weights than simple linear regression models. Therefore, we employ the coefficients obtained from ridge regression to initialize the shortcut connection layer, which provides better initial values in stabilizing and speeding up the training process. 

With $n$ training samples, let $\mathbf{X} \in \mathbb{R}^{n \times (2N-N_g-2)}$ and $\mathbf{y}^o_i \in \mathbb{R}^{n \times 1}$ denote the input data and the $i$-th dimension of the output data, respectively. Ridge regression finds the coefficient vector $\mathbf{w}_i \in \mathbb{R}^{(2N-N_g-2) \times 1}$ and bias $b_i \in \mathbb{R}$ by solving the problem:
\begin{align}
    \label{RR}
    \operatorname*{arg\,min}_{\mathbf{w}_i, b_i}\, \, L:= \|\mathbf{y}^o_i -(\mathbf{X} \mathbf{w}_i + b_i \mathbf{1}_n)\|_2^2 + \lambda \|\mathbf{w}_i\|_2^2\, , 
\end{align}
where $\mathbf{1}_n \in \mathbb{R}^{n}$ is the all-ones column vector. $\mathbf{w}_i^\top$ and $b_i$ will be used to initialized the $i$-th row of $\mathbf{W}_s$ and $\mathbf{b}_s$. The first term in the loss function \eqref{RR} calculates fitting errors of data pairs, while the second term shrinks the estimated coefficients. $\lambda$ is a tuning parameter that balances the relative strength of the least square error and the penalty term. When $\lambda = 0 $, it degrades to the least-square based linear regression.  

The closed-form solution to \eqref{RR} is derived as follows. We can rewrite the objective function $L$ as:
\begin{multline}
      L = \mathbf{w}_i^\top (\mathbf{X}^\top \mathbf{X} + \lambda \mathbf{I}) \mathbf{w}_i - 2\mathbf{w}_i^\top\mathbf{X}^\top\mathbf{y}^o_i + 2b_i\mathbf{w}_i^\top\mathbf{X}^\top\mathbf{1}_n \\
      -2b_i\mathbf{1}_n^\top\mathbf{y}^o_i + b_i^2\mathbf{1}_n^\top\mathbf{1}_n + \mathbf{y}^o_i{^\top} \mathbf{y}^o_i \, ,
\end{multline}
where $\mathbf{I} \in \mathbb{R}^{(2N-N_g-2) \times (2N-N_g-2)}$ is the identity matrix. The gradients of the convex objective function are  
\begin{align}
    \nabla_{\mathbf{w}_i} L &= -2(\mathbf{X}^\top\mathbf{y}^o_i - \mathbf{X}^\top \mathbf{X}\mathbf{w}_i - b_i\mathbf{X}^\top\mathbf{1}_n - \lambda \mathbf{w}_i) \, , \\
    \nabla_{b_i} L &= - 2 \mathbf{1}_n^\top\mathbf{y}^o_i + 2 \mathbf{1}_n^\top \mathbf{X} \mathbf{w}_i + 2b_i\mathbf{1}_n^\top\mathbf{1}_n \,.
\end{align}
Set $\nabla_{\mathbf{w}_i} L = \mathbf{0}$ and $\nabla_{b_i} L = 0$, we get:
\begin{align}
    \mathbf{w}_i &= (\mathbf{X}^\top \mathbf{X} + \lambda \mathbf{I})^{-1}(\mathbf{X}^\top \mathbf{y}^o_i -b_i \mathbf{X}^\top \mathbf{1}_n) \, , \label{weight_w}\\
    b_i &= \frac{1}{n} (\mathbf{1}_n^\top\mathbf{y}^o_i - \mathbf{1}_n^\top \mathbf{X} \mathbf{w}_i)\, \label{bias_b}.
\end{align}
After plugging \eqref{bias_b} into \eqref{weight_w}, we eliminate $b_i$ and obtain the closed-form solution of optimal $\mathbf{w}_i$ as:
\begin{align}
  \mathbf{w}_i = \left(\mathbf{X}^\top \mathbf{X} + \lambda \mathbf{I} - \mathbf{X}^\top \mathbf{H} \mathbf{X}\right)^{-1}\mathbf{X}^\top (\mathbf{I} -\mathbf{H}) \mathbf{y}^o_i
\end{align}
with $\mathbf{H} = \frac{1}{n} \mathbf{1}_n \mathbf{1}_n^\top \in \mathbb{R}^{n \times n}$.

\section{Numerical Results} \label{sec:Numerical Results}
With different datasets, the effectiveness of our proposed approaches is verified on the IEEE-30, IEEE-118, IEEE-300 \cite{Daniel2011}, SouthCarolina-500 \cite{Birchfield2017}, and PEGASE-1354 \cite{Josz2016} bus systems.  Detailed results and insights are presented in this section. 

\subsection{Simulation Setup}
\subsubsection{Test systems and datasets} We use synthetic and real-world data for a comprehensive evaluation. For load demands and power generation, the real-world data are provided by the global energy forecasting competition 2012 \cite{Hong2014} and PV plants installed in California \cite{nrel}, respectively. They are scaled to match the system capacity and avoid violations. In addition, the number of real-world data samples is insufficient for our simulated systems. Thus, we generate synthetic data as a supplement. For the IEEE-30 bus, the active power generation is modeled as multivariate Gaussian distributions. The ratio of standard deviation to mean value is 0.2, and the correlation coefficient between different PV buses is 0.2. In addition, the ratios of standard deviation to mean value are 0.1, 0.01, and 0.2 of the load demands on the IEEE-118, IEEE-300, and SouthCarolina-500 bus systems, respectively \cite{load_model}. The correlation coefficient between the same bus is 0.8 and 0.2 for different load buses. For the PEGASE-1354 bus system, the ratio of standard deviation to mean value is 0.1 for both active power generation and load demands. In addition, to consider various renewable energy uncertainties, we add six wind generations following the Weibull distribution and eight PV generations following the beta distribution. Besides, two generators may have power outages following the binomial distribution for the IEEE-118 bus system. We use IEEE-$118^*$ and PEGASE-$1354^*$ to denote those case studies. Finally, we conduct the PF analysis using the NR solver in \texttt{MATPOWER 7.0} to generate input-output data pairs \cite{Daniel2011}. The voltage angles are measured in radians.

\subsubsection{Methods for comparison} Based on the proposed framework shown in Fig.~\ref{fig:residual_block1}, we adopt four different initialization schemes, including random \cite{Kaiming2015}, data-driven PF model, linearized PF model, and Jacobian matrix model. We call them \emph{Random}, \emph{Data-driven}, \emph{Linearized PF}, and \emph{Jacobian}, respectively. The random initialization method serves as the baseline to show the benefits of our well-designed initialization methods. In addition, we also compare our work with some existing approaches.

\begin{itemize}    
    \item \emph{FC} \cite{Yang2020}: MLPs are used to learn the inverse AC-PF mappings.  Model-based initialization methods and loss functions are proposed. 
    
    \item \emph{TPBNN} \cite{Hu2021}: MLPs are applied to learn the inverse AC-PF equations, joined with an auxiliary task to rebuild the forward PF mapping. The training loss function consists of both estimation error and reconstruction error. 
        
    \item \emph{ResNet} \cite{He2016}: ResNets consist of stacked residual building blocks. Each block replaces the identity mapping (cf. Fig.~\ref{fig:residual_block}) with a fully connected linear layer to improve the generalization capability. The output layer is also linear because voltage angles can be negative. 

    \item \emph{LPF} \cite{Yu2016}: AC-PF equations are linearized around the operating point by the first-order Taylor expansion. 

    \item \emph{SML} \cite{Fu2020}: The stochastic response surface method adopts the polynomial chaos expansion to model the input-output random variables relationship. 
    
    \item \emph{KNN}: K nearest neighbor algorithm predicts the corresponding voltage phasors of new power injection samples based on their feature similarity measure to the training data points. It belongs to non-linear regressors and has various applications in power systems, e.g., state estimation, load forecasting, and fault detection \cite{Miraftabzadeh2021}. 
    
    \item \emph{RR}: We employ ridge regression to learn the inverse AC-PF mappings (see eq. \eqref{RR}).  
    
    \item \emph{SVR} \cite{Yu2017}: Support vector regression can approximate the non-linear AC-PF equations with the kernel trick. 

    \item \emph{QMC} \cite{Xu2017}: Uses the low discrepancy sequences to achieve a faster convergence rate than traditional MCS, which uses simple random sampling. It can reduce the computational burden by decreasing the number of samples without sacrificing accuracy. 
    
\end{itemize}

\subsubsection{Details of training} 
Adam optimizer with mini-batch size 32 is used for training. Table \ref{NN_str} shows the network structure and the dataset size of five benchmark systems. The validation dataset is used to tune the hyperparameters. In addition, the mean square error (MSE) is adopted as the loss function. The training process will stop when the validation loss has stabilized with no further improvements \cite{Prechelt2012}. We train and test all models five times to alleviate the randomness. 


\begin{table}
\caption{Hyperparameters of the NN structures. The second column indicates the number of neurons in each layer. The last column shows the size of each dataset.}
\label{NN_str}
\centering
\begin{tabular}{c|c|c}\hline 
Cases & Structure & [Training, Validation, Testing]\\\hline
30 & [53 100 100 53] & [12k, 4k, 4k]  \\\hline
118 & [181 300 300 181] & [20k, 5k, 5k] \\\hline
300 & [530 200 200 200 530] & [20k, 5k, 5k] \\\hline
500 & [909 300 300 300 300 909] & [28k, 6k, 6k] \\\hline
1354 & [2447 300 300 300 300 2447] & [34k, 8k, 8k] \\\hline
\end{tabular}
\end{table}
\subsection{Model Evaluation Criteria}
\subsubsection{Average root mean square error (ARMSE)} Let $\mathbf{O}$, $\hat{\mathbf{O}} \in \mathbb{R}^{M \times D}$ denote the matrices containing the true and estimate values. $M$ and $D$ are the numbers of samples and output dimensions. $O_{i,j}$ and $\hat{O}_{i,j}$ are the $(i,j)$-th element of $\mathbf{O}$ and $\hat{\mathbf{O}}$, respectively. The ARMSE can be calculated by:
\begin{align}
    \text{ARMSE} =\frac{1}{D} \sum_{j=1}^D \sqrt{\frac{1}{M}\sum_{i=1}^M (\hat{O}_{i,j} - O_{i,j})^2}\, .
\end{align}

\subsubsection{Mean absolute percentage error (MAPE)} The MAPE for the $j$-th element of the outputs is given as:
\begin{align}
    \text{MAPE} = \frac{1}{M} \sum_{i=1}^M \left| \frac{\hat{O}_{i,j} - O_{i,j}}{O_{i,j}}\right| \times 100\%\, .
\end{align}

\subsubsection{Average Wasserstein distance (AWD)} Wasserstein distance measures the similarity between two probability distributions in the same metric space. Let $\rho_j$ and $\hat{\rho}_j$ be the distributions of the $j$-th column of $\mathbf{O}$ and $\hat{\mathbf{O}}$, respectively. The first-order Wasserstein distance loss $l_{\mathrm{wd}}$ between the two distributions is calculated by:
\begin{align}
    l_{\mathrm{wd}}(\hat{\rho}_j, \rho_j) = \underset{\gamma \in \Gamma (\hat{\rho}_j, \rho_j)}{\text{inf}} \int_{\mathbb{R} \times \mathbb{R}}|\hat{\rho}_j - \rho_j|  \text{d} \gamma(\hat{\rho}_j, \rho_j)\, ,
\end{align}
where $\Gamma (\hat{\rho}_j, \rho_j)$ denotes the set of all measures on $\mathbb{R} \times \mathbb{R}$ whose marginal distributions are $\hat{\rho}_j$ and $\rho_j$ on the first and second factors, respectively. Then, the average loss of all output dimensions is:
\begin{align}
    \text{AWD} = \frac{1}{D}\sum_{j=1}^{D} l_{\mathrm{wd}}(\hat{\rho}_j, \rho_j).
\end{align}

\subsection{Power Flow Analysis Results}
Table~\ref{rmse} shows our proposed approaches outperform the other methods in predicting the voltage magnitudes and angles for the PF analysis. Besides, we have the following observations:

\begin{itemize}
    \item Classical non-linear regressors in the machine learning field, including KNN, RR, and SVR, have significant estimation errors in medium- and large-scale bus systems. These solvers cannot effectively extract the data features and recover the complicated mapping relationship in the inverse AC-PF equations.
    
    \item The NN-based approaches achieve more accurate predictions than the linear LPF method, the non-linear SML method, and classical non-linear regressors, especially in the voltage angle estimates. 
    
    \item The performance of the FC method is comparable to that of the ResNet and Random approaches. It shows that only using the residual framework does not have apparent advantages over the vanilla MLP structure. 
    
    \item Three designed initialization schemes seem straightforward but significantly outperform the FC and random initialization methods in improving the accuracy of voltage phasor estimates.
    
\end{itemize} 

\begin{figure}[t]
    \centering
    \includegraphics[width = 0.45\textwidth]{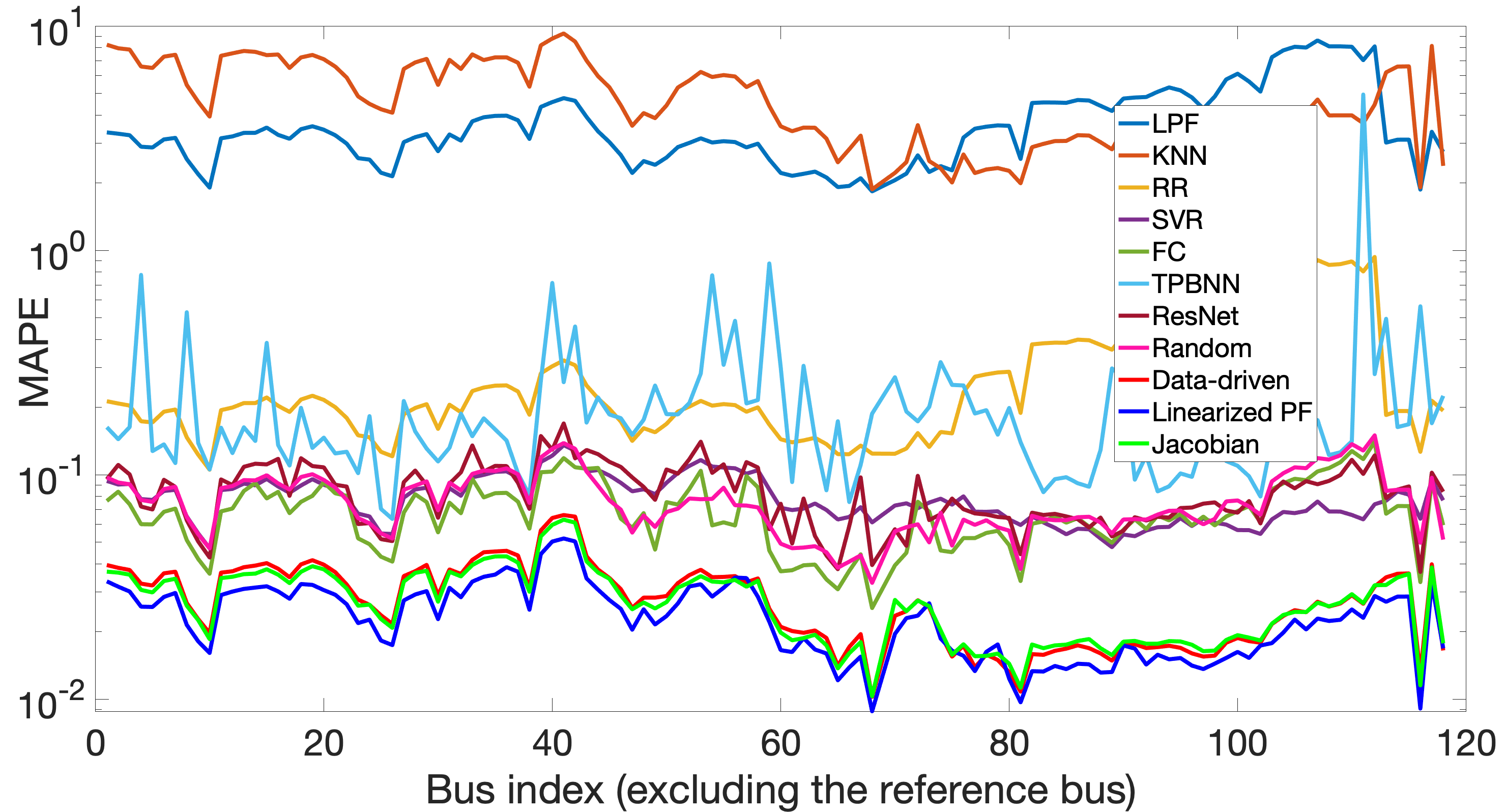}
    \caption{MAPEs of the voltage angles for the IEEE-118 bus system.}
    \label{fig:MAPE_118_ang}
\end{figure}

\begin{figure}[t]
    \centering
    \includegraphics[width = 0.45\textwidth]{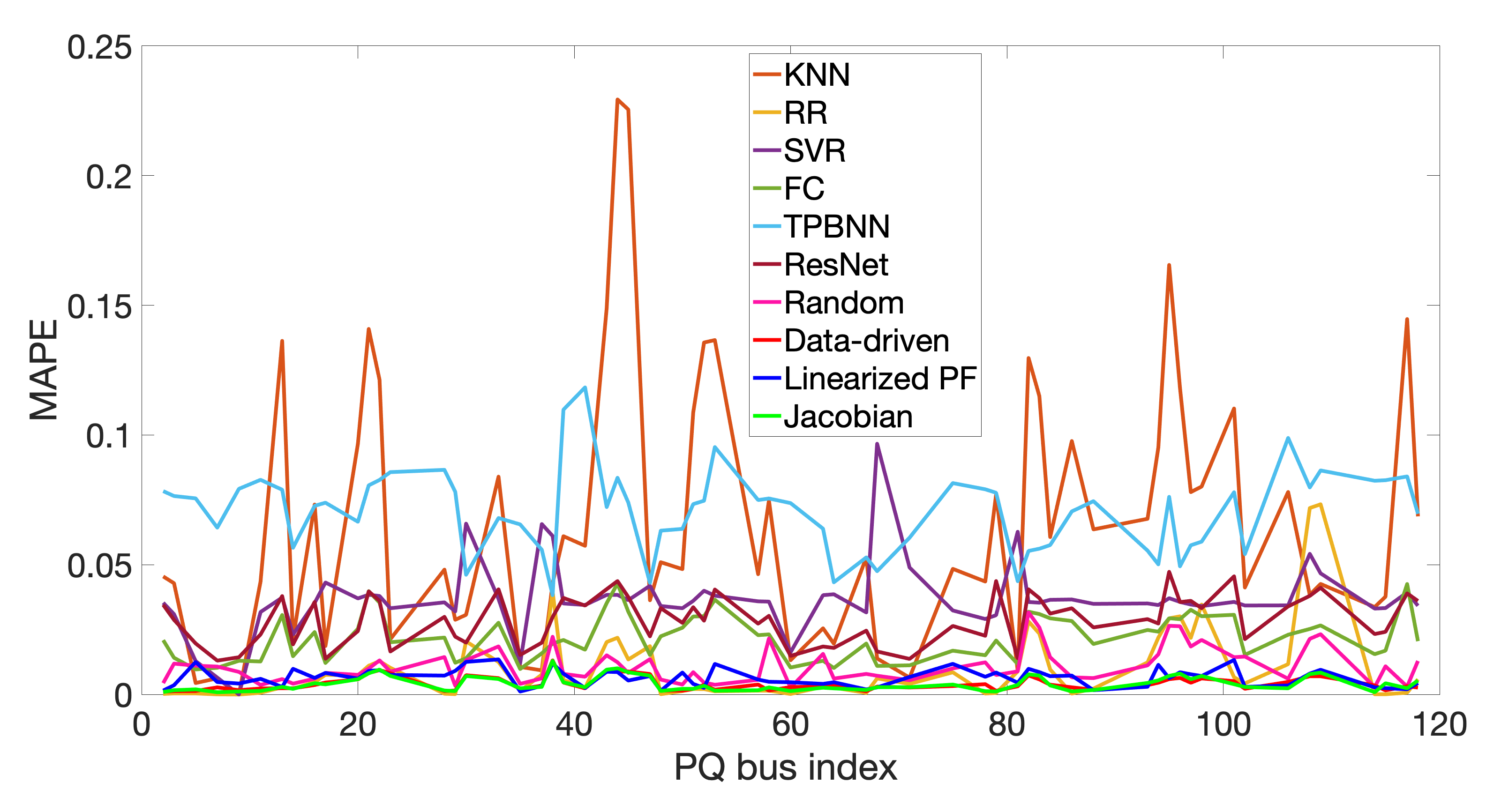}
    \caption{MAPEs of the voltage magnitudes for the IEEE-118 bus system.}
    \label{fig:MAPE_118_mag}
\end{figure}

In addition, Fig.~\ref{fig:MAPE_118_ang} and Fig.~\ref{fig:MAPE_118_mag} show the MAPEs of voltage angles and magnitudes on the IEEE-118 bus system. The average MAPEs of voltage angles and magnitudes of the data-driven method are 0.028\% and 0.0042\%. Similarly, 0.023\% and 0.0065\% for the Linearized PF method, and 0.023\% and 0.0042\% for the Jacobian method. 

Table~\ref{rmse_branch} shows the errors of branch flow calculations. Data-driven, Linearized PF and Jacobian schemes achieve the best performance among all the NN-based solvers. However, on the IEEE-300 and PEGASE-1354 bus systems, the LPF and RR methods obtain accurate branch flow calculations despite their poor performance in voltage phasor estimates. The reason is as follows. The relationship between voltage angles and voltage angle differences is not bijective. Branch flows between two buses are more related to voltage angle differences than voltage angles. However, the outputs of our proposed models are voltage angles instead of voltage angle differences. Therefore, our methods may predict $\theta_i$ and $\theta_j$ accurately, while there is no guarantee of a similar level of accuracy for $\theta_{ij}$ estimate. This property further affects the calculation accuracy of branch flows \cite{Kejun2022}. 

\begin{table*}
\caption{ARMSEs of the voltage phasor calculations for different cases ($10^{-4}$).}
\label{rmse}
\centering
\scalebox{0.9}{
\begin{tabular}{c|c|c|c|c|c|c|c|c|c|c|c|c|c}\hline 
Cases & Voltage phasor & LPF & SML & KNN & RR & SVR & FC & TPBNN & ResNet & Random & Data-driven & Linearized PF
& Jacobian \\\hline

\multirow{2}{*}{30} & angle & 188.40 & 479.04 & 110.45 & 3.66 & 1.51  & 2.96 & 6.29 & 2.17 & 2.06 & 1.43 & \textbf{1.35} &1.94\\
& magnitude & 13.72 & 41.33 & 25.21 & 1.78  & 1.27  & 2.17 & 5.20 & 1.44 & 1.23 & 1.08 & \textbf{1.02} &1.16  \\ \hline

\multirow{2}{*}{118} & angle & 289.31 & 1661.08 & 342.04 & 24.33  & 5.61 & 7.36 & 61.37 & 8.14  & 7.14 & 2.70 & \textbf{2.46} & 2.72 \\
& magnitude & 11.93 & 20.95 & 8.39 & 1.43 & 4.36 & 4.07 & 9.84 & 5.09 & 2.93 & \textbf{0.79} & 1.24 & 0.85\\ \hline

\multirow{2}{*}{118$^*$} & angle & 87.67 & 1754.87 & 385.79 & 17.44 & 5.76 & 7.50 & 15.00  & 10.82 & 5.54 & 2.55 & 2.71 & \textbf{2.48} \\

& magnitude & 5.03 & 17.54 & 8.57 & 1.55 & 4.40 & 5.30 & 9.94 & 6.80 & 2.10 & \textbf{0.76} & 0.80 & 0.77 \\ \hline

\multirow{2}{*}{300} & angle & 302.38 & 2476.92 & 579.51 & 108.55 & 324.10 & 21.71 & 32.66 & 35.11  & 23.59 & 12.87 & 7.80 & \textbf{6.96} \\
& magnitude & 27.94 & 24.76 & 9.73 & 11.23  & 9.63 & 5.58 & 8.20  & 8.81 & 10.65 & 2.42 & 2.18 & \textbf{1.94}\\ \hline

\multirow{2}{*}{500} & angle & 309.28 & 4284.63 & 401.38 & 259.15 & 239.32 & 16.18 & 798.32 & 13.40 & 7.83 & 8.95  & \textbf{6.93} & 8.26 \\ 
& magnitude & 58.46 & 357.44 & 46.81 & 46.68 & 238.68 & 10.91 & 15.13 & 8.48 & 3.35 & 3.67 & \textbf{3.31} & 3.43 \\ \hline

\multirow{2}{*}{1354} & angle & 207.19 & 6145.52 & 1189.77 & 164.72 & 2973.84 & 32.20 & - & 30.84  & 13.79 & 12.77 & 9.13 & \textbf{8.78} \\
& magnitude & 7.82 & 48.95 & 16.78 & 6.59  & 60.23 & 10.23 & -  & 10.11 & 15.53 & 4.34 & \textbf{3.53} & 3.54\\ \hline

\multirow{2}{*}{1354$^*$} & angle & 80.47  & 4707.09 & 374.57 & 65.61 & 2059.95  & 18.88 & -& 29.85 & 21.15 & 6.74 & \textbf{5.11} & 5.20 \\
& magnitude & 3.84 & 45.73 & 11.00 & 3.58 & 36.48 & 10.67 & - & 12.05 & 12.31 & 3.50 & 3.31 & \textbf{3.13} \\  \hline

\end{tabular}
}
\end{table*}

\begin{table*}
\caption{ARMSEs of the branch flow calculations for different cases}
\label{rmse_branch}
\centering
\scalebox{0.94}{
\begin{tabular}{c|c|c|c|c|c|c|c|c|c|c|c|c|c}\hline 
Cases & Branch flow & LPF & SML & KNN & RR & SVR & FC & TPBNN & ResNet & Random & Data-driven & Linearized PF & Jacobian   \\\hline

\multirow{2}{*}{30} & active & 3.998 & 5.049 & 1.329 & 0.038  & 0.080 & 0.108  & 0.618 & 0.057 & 0.029 & \textbf{0.028}
& 0.031 & \textbf{0.028}\\
& reactive & 1.064 & 1.774 & 0.660 & 0.048 & 0.088 & 0.103 & 0.457 & 0.056 & 0.036 & 0.034 & 0.041 & \textbf{0.033}\\ \hline

\multirow{2}{*}{118} & active & 2.836 & 19.944 & 4.472 & 0.282 &  0.294 & 0.391 & 3.585 & 0.745 & 0.255 & \textbf{0.071} & 0.105 & 0.077 \\
& reactive & 1.341 & 4.579 & 1.063 & 0.153 & 0.643 & 0.321 & 1.654 & 0.440 & 0.221  &  0.067 & 0.108 & \textbf{0.065}\\ \hline

\multirow{2}{*}{118$^*$} & active & 0.678 & 21.288 & 5.004 & 0.156 & 0.347 & 0.588 & 4.235 & 0.924 & 0.204 & 0.089 & \textbf{0.070} &  0.078    \\ 
& reactive & 0.710 & 4.447 & 1.173 & 0.191 & 0.611 & 0.403 & 1.719 &  0.529 & 0.172 & 0.075 & 0.065  & \textbf{0.063} \\ \hline

\multirow{2}{*}{300} & active & 0.930 & 14.010 & 2.889 & \textbf{0.209} & 2.212 & 1.718 & 2.430 & 4.526 & 5.146 & 0.496  &  0.557 & 0.508 \\
& reactive & 2.314 & 3.566 & 0.911 & 0.686 & 1.239 & 0.924 & 2.080  & 1.817 &  2.869 & 0.471 & 0.452 & \textbf{0.427} \\ \hline

\multirow{2}{*}{500} & active & 1.658 & 37.855 & 7.821 & 1.427 & 9.974 & 4.394 & 17.896 & 2.834 & 0.970 & 0.756 & 0.767 & \textbf{0.741} \\
& reactive & 3.843 & 18.909 & 3.899 & 3.095 & 21.054 & 3.015 & 81.021 & 2.091 & 1.205 & 0.909 & \textbf{0.798} & 0.859 \\ \hline

\multirow{2}{*}{1354} & active & \textbf{1.103} & 62.524 & 18.677 & 1.131 & 42.54 & 16.164 & - & 15.843 & 11.94 & 7.066 & 5.808 & 5.922 \\
& reactive & 2.197 & 15.189 & 4.201 & \textbf{1.857} & 20.48 & 7.800 & - & 8.189 & 19.65 & 6.841 & 5.395 & 5.406 \\ \hline

\multirow{2}{*}{1354$^*$} & active & 0.513 & 44.810 & 8.420 & \textbf{0.420} & 50.087 & 7.660 & - & 19.056 & 34.828 & 3.986 & 4.055 & 3.923 
\\ 
& reactive & 0.996 & 9.812 & 2.252 & \textbf{0.812} & 21.279 & 10.821 & - &  14.658 & 23.948 & 5.626 & 5.033 & 5.305
\\ \hline
\end{tabular}
}
\end{table*}

\subsection{Comparison Results of Probabilistic Characteristics} 
\subsubsection{Wasserstein distance comparison} The Wasserstein distance measures the minimum effort in transforming the probability mass from one distribution to the other. It quantifies the distribution difference between the model outputs and the targets \cite{wgan}. Table~\ref{AWD} and Table~\ref{AWD_BF} show the AWD of voltage phasor and branch flow distributions. Two physics-guided initialization schemes achieve the best performance among all the methods. 

\begin{table*}
\caption{AWDs of the voltage phasor distributions for different cases ($10^{-4}$).}
\label{AWD}
\centering
\scalebox{0.92}{
\begin{tabular}{c|c|c|c|c|c|c|c|c|c|c|c|c|c}\hline 

Cases & Voltage phasor & LPF & SML & KNN & RR & SVR & FC & TPBNN & ResNet & Random & Data-driven & Linearized PF & Jacobian  \\\hline
\multirow{2}{*}{30} & angle & 152.99 & 33.50 & 45.84 &2.07 & 0.67 & 0.86 & 2.03 & 0.70 & 0.76 & 0.53 & \textbf{0.49} & 0.70 \\
& magnitude & 10.73 & 4.34 & 13.70 & 0.59 & 0.57 & 0.51 & 1.99 & 0.35 & 0.37 & 0.28 & \textbf{0.25} & 0.32  \\ \hline

\multirow{2}{*}{118} & angle & 257.96 & 66.79 & 158.91 & 12.7  & 2.64 & 1.96 & 10.43 & 2.22  & 2.48 & 0.92 & \textbf{0.86} & 0.91 \\
& magnitude & 10.17 & 0.97 & 3.64 & 0.58  & 2.26 & 0.94 & 2.98 & 1.32 & 0.74 & 0.22 & 0.52 & \textbf{0.20} \\ \hline

\multirow{2}{*}{118$^*$}  & angle & 78.95 & 109.10 & 175.09 & 11.12 & 2.65 & 2.04 & 3.55 & 2.70 & 2.03 & 0.96 & 0.98 & \textbf{0.89} \\
& magnitude & 3.88 & 1.21 & 4.00 & 0.46 & 2.15 & 1.09 & 2.80 & 1.41 & 0.45 & 0.24
& \textbf{0.20} & 0.22  \\ \hline

\multirow{2}{*}{300} & angle & 266.14 & 212.63 & 144.58 & 69.72 & 84.50 & 4.57 & 5.81 & 7.07  & 4.75 & 3.49 & 2.36 &\textbf{2.22}  \\
& magnitude & 22.44 & 3.08 & 3.46 & 4.08 & 4.23 & 1.29 & 2.53  & 1.78 & 3.14 & 0.60 & 0.62 & \textbf{0.52}  \\ \hline

\multirow{2}{*}{500} & angle & 167.55 & 93.60 & 199.68 & 160.44 & 191.47 & 3.26 & 59.16 & 2.82 & 1.75 & 2.44 & \textbf{1.71} & 1.82\\
& magnitude & 35.30 & 10.81 & 19.49 & 29.06 & 221.55 & 1.62 & 3.24 & 1.22 & 1.05 & 1.19 & 0.97 & \textbf{0.89}  \\ \hline

\multirow{2}{*}{1354} & angle & 108.80 & 85.46 &659.17 & 101.01  & 2364.51 & 5.79 & - & 5.66  & 5.22 & 3.92 & 2.79 &\textbf{2.72}  \\
& magnitude & 4.75 & 1.05 & 9.23 & 2.40 & 55.00 & 2.26 & -  & 2.75 & 5.32 & 1.53  & \textbf{1.04} & 1.09  \\ \hline

\multirow{2}{*}{1354$^*$} & angle & 47.65 & 59.32 & 195.80 & 42.88 & 1621.93 & 3.76 & - & 4.78 & 3.76 & 2.11 & 1.44 & \textbf{1.39}  \\
& magnitude & 2.36 & \textbf{0.60} & 4.32 & 1.69 & 31.44 & 1.59 & - & 2.50 & 3.21 & 1.40 & 1.27 & 1.26 \\ \hline

\end{tabular}
}
\end{table*}

\begin{table*}
\caption{AWDs of the branch flow distributions for different cases.}
\label{AWD_BF}
\centering
\scalebox{0.95}{
\begin{tabular}{c|c|c|c|c|c|c|c|c|c|c|c|c|c}\hline 

Cases & Branch flow & LPF & SML & KNN & RR & SVR & FC & TPBNN & ResNet & Random & Data-driven & Linearized PF & Jacobian   \\\hline
\multirow{2}{*}{30} & active & 3.293 & 0.382 & 0.625 & 0.019  & 0.022 & 0.024  & 0.246 & 0.019 & 0.013 & 0.016
& 0.018 & \textbf{0.012}\\
& reactive & 0.871 & 0.166 & 0.319 & 0.014 & 0.033 & 0.025 & 0.194 & 0.014 & 0.012 & 0.014 & 0.025 & \textbf{0.010}\\ \hline

\multirow{2}{*}{118} & active & 2.420 & 1.291 & 1.988 & 0.145 & 0.103 & 0.121 & 1.740 & 0.213 & 0.147 & 0.045 & 0.080 & \textbf{0.039}   \\
& reactive & 1.144 & 0.326 & 0.456 & 0.074 & 0.404 & 0.117 & 0.715 & 0.160 & 0.115  &  0.039 & 0.085 & \textbf{0.030}\\ \hline

\multirow{2}{*}{118$^*$} & active & 0.560 &  2.059 & 2.414 & 0.094 & 0.102 & 0.121 & 2.090 & 0.195 & 0.092 & 0.063 & \textbf{0.038} & 0.045 \\
& reactive & 0.556 & 0.441 & 0.548 & 0.093 & 0.333 & 0.102 & 0.760 & 0.141 & 0.069 & 0.048 &
\textbf{0.035} & 0.036 \\ \hline

\multirow{2}{*}{300} & active & 0.781 & 1.563 & 1.010 & \textbf{0.134} & 1.069 & 0.644 & 1.043 & 2.066 & 2.894 & 0.249 & 0.268 & 0.256\\
& reactive & 1.880 & 0.433 & 0.296 & 0.325 & 0.754 & 0.396 & 1.049  & 0.907 &  2.073  & 0.347 & \textbf{0.289} & 0.296 \\ \hline

\multirow{2}{*}{500} & active & 0.924 & 0.690 & 4.026 & 0.860 & 7.330 & 1.273 & 6.772 & 0.922 & 0.540 & 0.395 & \textbf{0.356} & \textbf{0.356} \\
& reactive &  2.278 & \textbf{0.556} & 1.869 & 1.935 & 18.031 & 1.176 & 7.537 & 0.803 & 0.922 & 0.681 & 0.569 & 0.630\\ \hline

\multirow{2}{*}{1354} & active & \textbf{0.611} & 1.162 & 10.157 & 0.629 & 28.185 & 7.089 & - & 6.815 & 9.806 & 4.493 & 3.077 & 3.148\\
& reactive & 1.274 & 5.949 & 6.703 & \textbf{0.890} & 20.103 & 4.523 & -  & 5.277 &  14.079  & 5.105 & 3.497 & 3.589 \\ \hline

\multirow{2}{*}{1354$^*$}  & active & 0.303 & 0.611 & 4.043 & \textbf{0.257} & 28.666 & 3.098 & - & 11.106 & 25.222 & 3.140 & 3.132 & 3.092
\\ 
& reactive & 0.589 & \textbf{0.142} & 1.002 & 0.441 & 18.318 & 2.600 & - & 7.301 & 18.161 & 5.101 & 4.412 & 4.793
\\ \hline
\end{tabular}
}
\end{table*}

\begin{figure}[t]
    \centering
    \includegraphics[width = 0.45\textwidth]{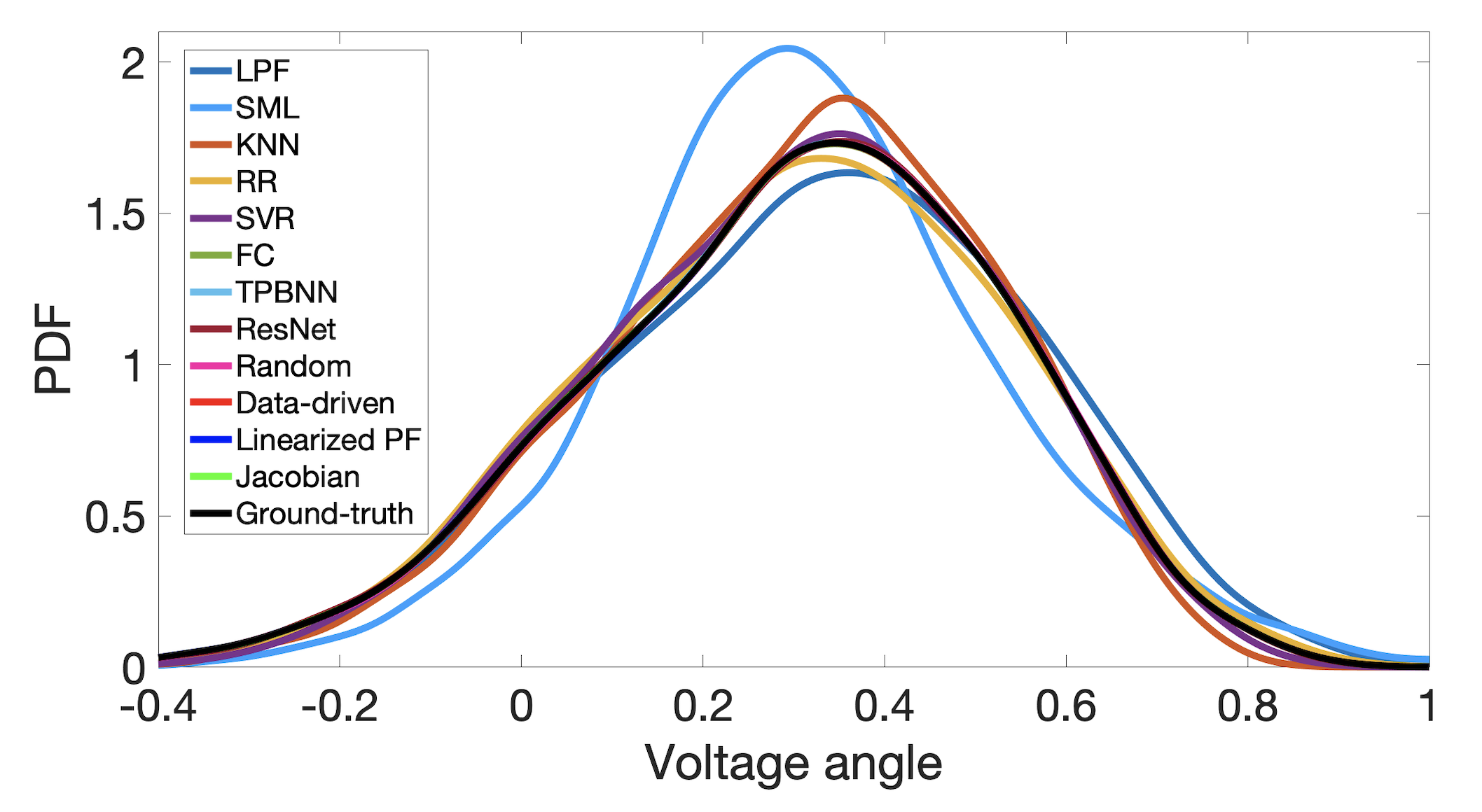}
    \caption{PDFs of the voltage angle of bus 1 for the IEEE-300 bus system.}
    \label{fig:pdf_ang118}
\end{figure}
\begin{figure}[t]
    \centering
    \includegraphics[width = 0.45\textwidth]{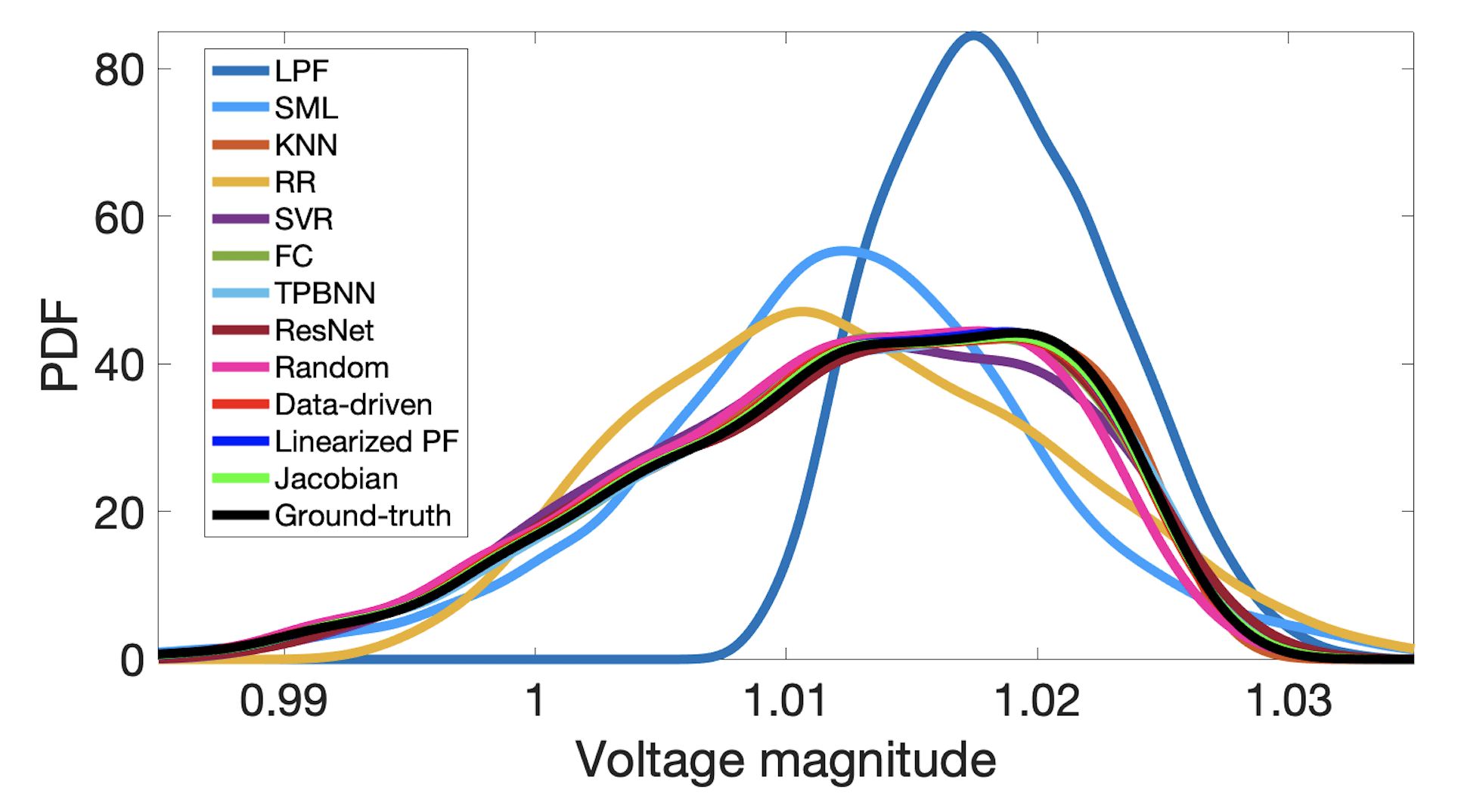}
    \caption{PDFs of the voltage magnitude of bus 16 for the IEEE-300 bus system.}
    \label{fig:pdf_mag118}
\end{figure}

\begin{figure}[t]
    \centering
    \includegraphics[width = 0.45\textwidth]{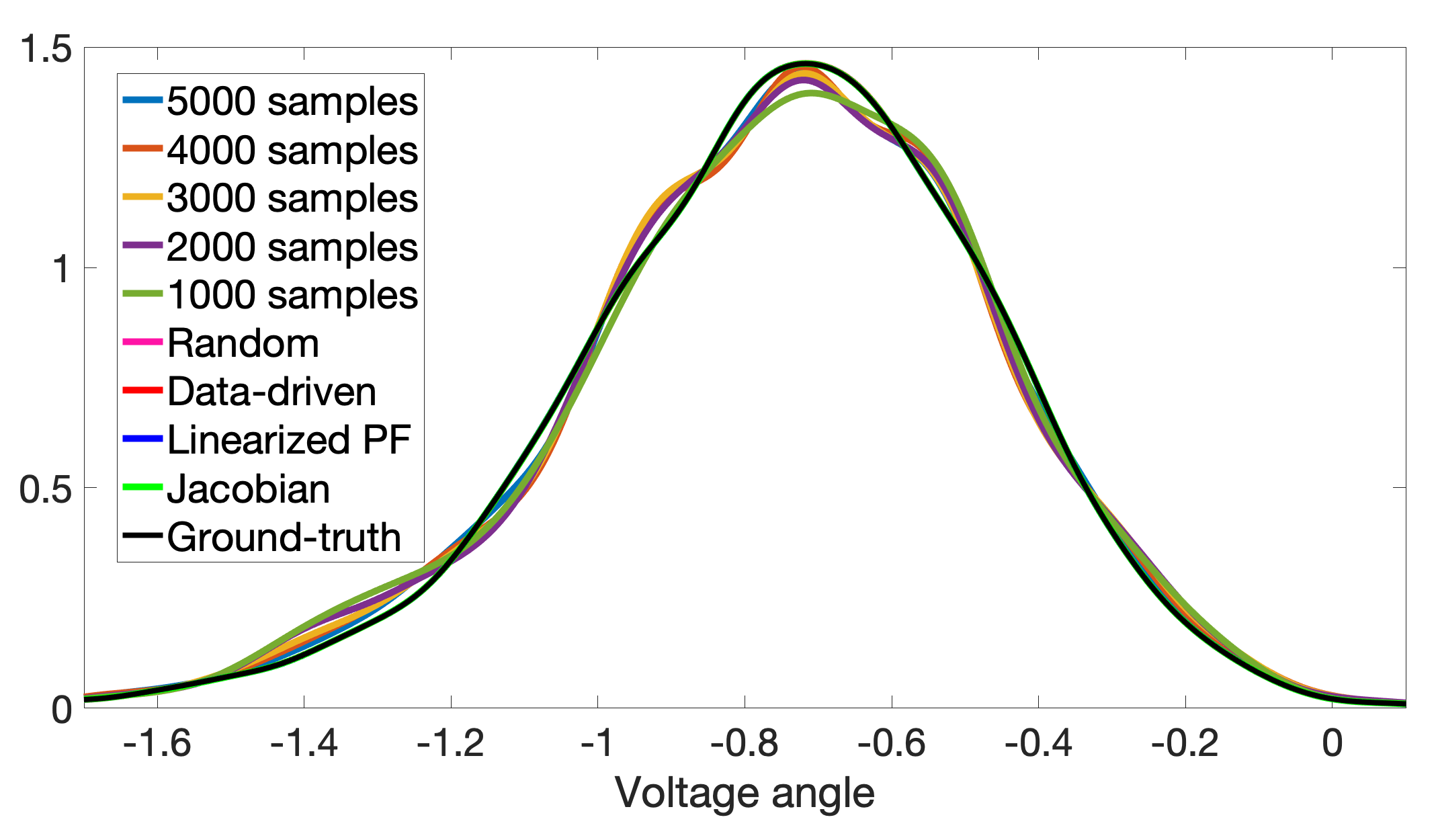}
    \caption{PDFs of the voltage angle of bus 1 for the SouthCarolina-500 bus system.}
    \label{fig:pdf_angQMC}
\end{figure}
\begin{figure}[t]
    \centering
    \includegraphics[width = 0.45\textwidth]{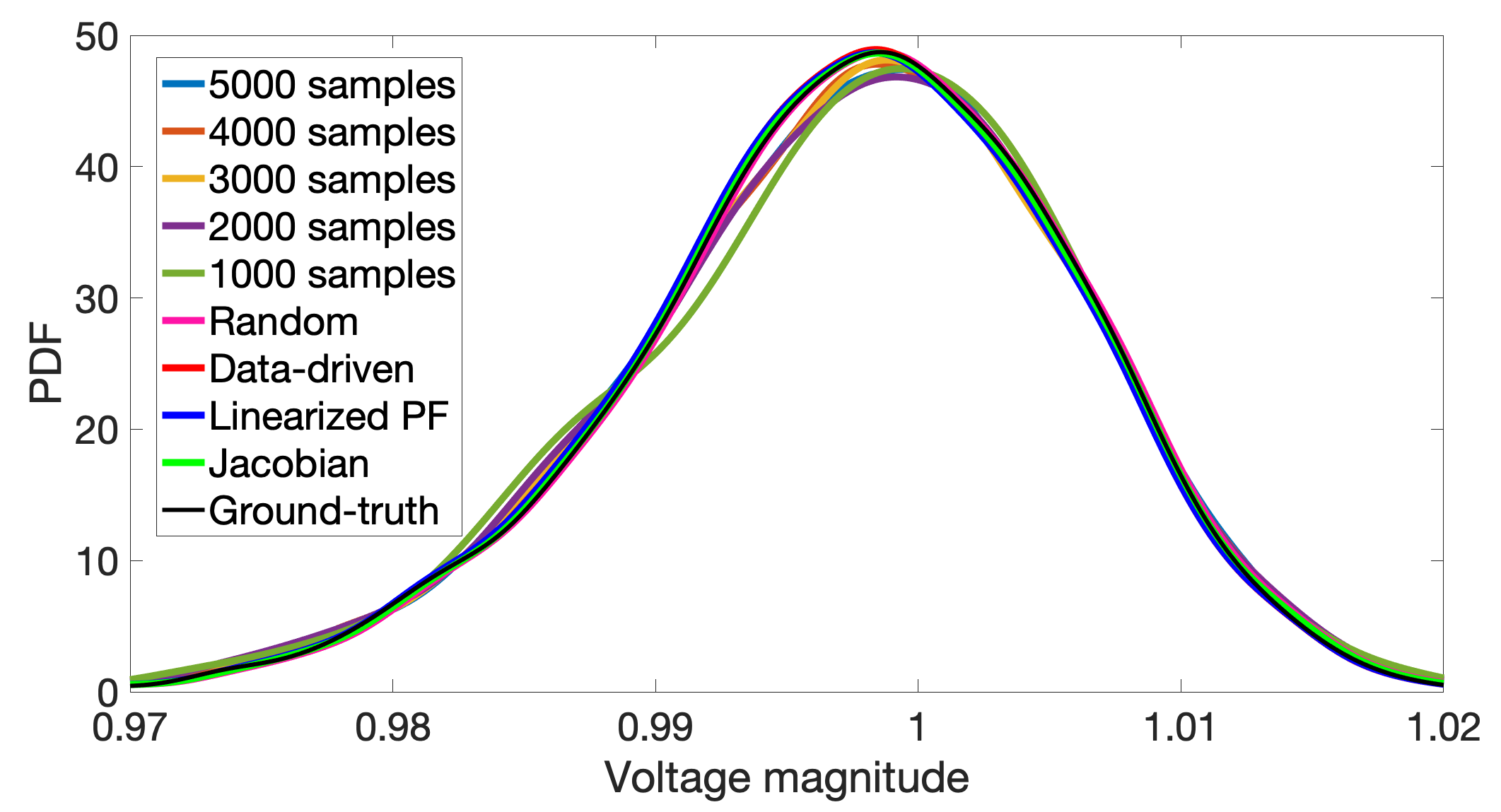}
    \caption{PDFs of the voltage magnitude of bus 1 for the SouthCarolina-500 bus system.}
    \label{fig:pdf_magQMC}
\end{figure}

\begin{figure}[t]
    \centering
    \includegraphics[width = 0.45\textwidth]{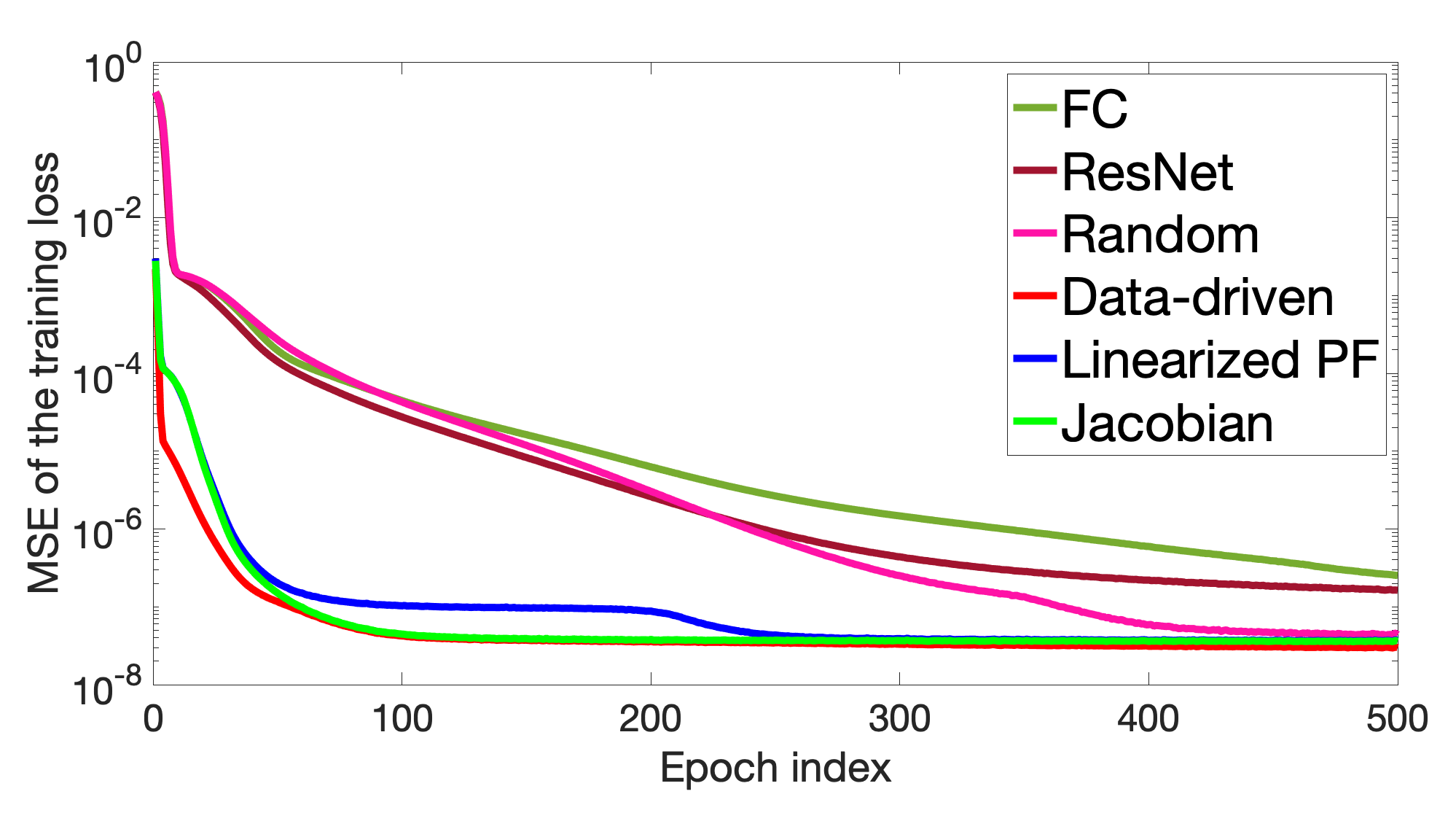}
    \caption{The training loss evolution (starting from when the first epoch's training is done) for the IEEE-30 bus system.}
    \label{fig:mse_loss30}
\end{figure}

\begin{figure}[t]
    \centering
    \includegraphics[width = 0.45\textwidth]{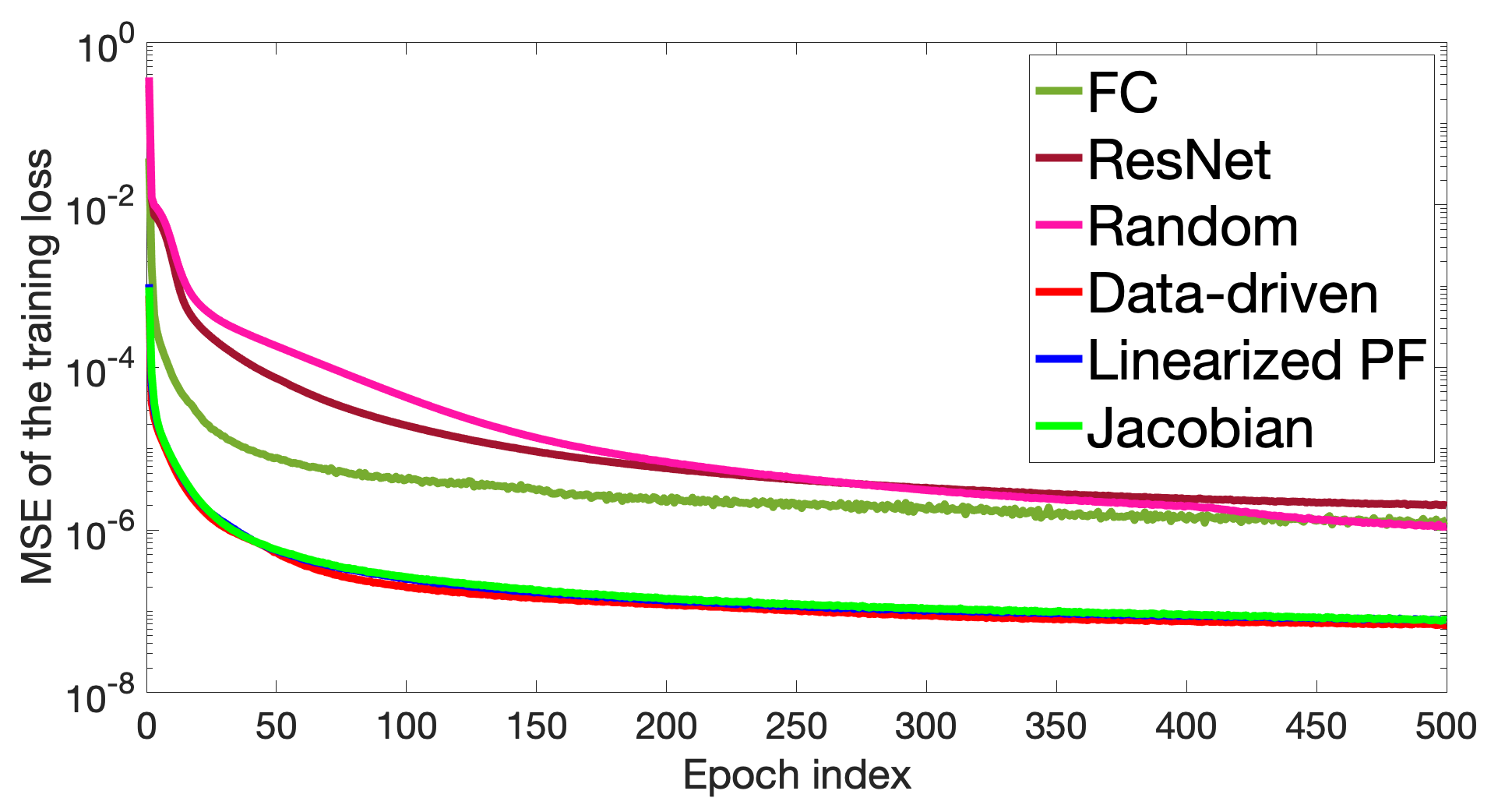}
    \caption{The training loss evolution (starting from when the first epoch's training is done) for the IEEE-118 bus system.}
    \label{fig:mse_loss118}
\end{figure}

\begin{figure}[t]
    \centering
    \subfloat[Initial weights \label{fig:imgshow_initial_random}]{%
    \includegraphics[scale=0.38]{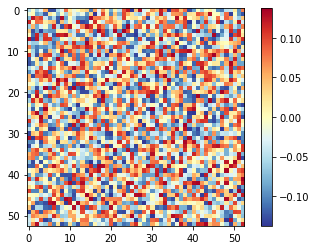}}
    \hfill
    \subfloat[Weights after training 500 epochs 
    \label{fig:imgshow_later_random}]{%
    \includegraphics[scale=0.38]{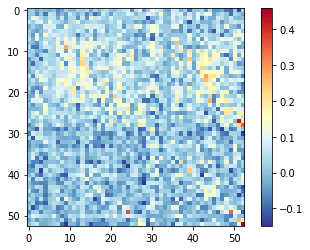}}
    \caption{Weights of the shortcut connection linear layer of the Random method for the IEEE-30 bus system.}
\end{figure}

\begin{figure}[t]
    \centering
    \subfloat[Initial weights \label{fig:imgshow_initial}]{%
    \includegraphics[scale=0.4]{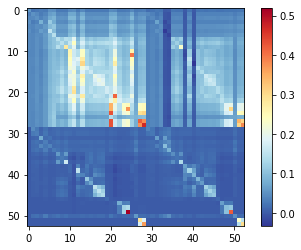}}
    \hfill
    \subfloat[Weights after training 500 epochs 
    \label{fig:imgshow_later}]{%
    \includegraphics[scale=0.4]{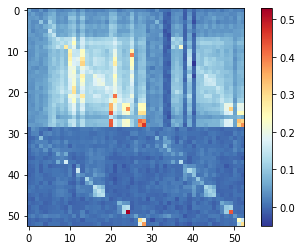}}
    \caption{Weights of the shortcut connection linear layer of the data-driven method for the IEEE-30 bus system.}
    \label{fig:imgshow_all}
\end{figure}

\subsubsection{PDF estimates accuracy comparison} Fig.~\ref{fig:pdf_ang118} and Fig.~\ref{fig:pdf_mag118} show the estimated and target PDFs. The voltage magnitude of bus 16 has the largest standard deviation value, which means its voltage magnitude values are spread out over a broader range. Therefore, its PDF estimate accuracy can be a good indicator for comparing different approaches. As shown in Fig.~\ref{fig:pdf_mag118}, the estimated PDFs obtained by our proposed approaches are almost close to the ground-truth PDF. In addition, it is worth pointing out that the PDF estimates obtained by the LPF, SML, and RR methods have significant errors. 

Furthermore, the accuracy of MCS can be improved by feeding more data samples, thus providing more precise PDF estimates. Employing a fixed number of iterations or establishing a threshold for the variance is practical for determining the stopping criteria of the MCS \cite{PRUSTY20171286}. Table \ref{sample} shows variance coefficients of voltage phasors using different sampling numbers of the MCS method. Nonetheless, it is crucial to acknowledge that the variance coefficient primarily emphasizes the mean value and variance. Achieving a precise PDF estimation might require a larger dataset, extending beyond precise mean value and variance estimations. Consequently, even if the coefficient of variation converges, the PDF estimate could still lack the desired level of accuracy.

In addition, Fig.~\ref{fig:pdf_angQMC} and Fig.~\ref{fig:pdf_magQMC} show the estimated PDFs obtained from the QMC with different sampling numbers. The QMC method aims to reduce sampling numbers while guaranteeing accurate PDF estimates. However, with the decrease in sampling numbers, the estimated PDFs are also slowly driving far away from the ground-truth PDF. Therefore, to guarantee an accurate PDF estimate, we cannot significantly reduce the sampling points. Besides, the computational complexity of the NR solver with $k$ iterations is $\mathcal{O}(k \times N^{1.4})$ for each sample \cite{Alvarado1976}. In a word, the improved sampling algorithm QMC cannot significantly reduce the total computational time due to the necessity of enough samples and the usage of the NR solver. In addition, we notice that our proposed approaches can achieve accurate PDF estimation. 

\begin{table}
\caption{The variance coefficients of voltage phasors using different sampling numbers of the MCS for different cases.}
\label{sample}
\centering
\scalebox{0.9}{
\begin{tabular}{|c|c|c|c|c|}\hline 
Cases & 2000 samples & 3000 samples & 4000 samples  & 5000 samples  \\\hline

\multirow{1}{*}{30} & 0.079 & 0.063 & 0.055 & 0.049 \\ \hline

\multirow{1}{*}{118} & 0.005 & 0.004 & 0.004 & 0.003 \\  \hline

\multirow{1}{*}{300}  & 1.805 & 12.914 & 2.280 & 1.240 \\  \hline

\multirow{1}{*}{500}  & 0.033 & 0.026 & 0.022 & 0.020  \\  \hline

\multirow{1}{*}{1354} & 5.783 & 20.016 & 2.718 & 5.971   \\  \hline

\end{tabular}
}
\end{table}

\subsection{Risk assessment}
The performance of the proposed approaches is evaluated based on risk assessment, which is critical to power system operation. Tables~\ref{risk} and~\ref{risk_bf} show the probabilities of exceeding the operational limit of the voltage magnitude and apparent branch flow on the SouthCarolina-500 bus system  \cite{Xu2020_risk}. Compared with the MCS method, our proposed approaches have achieved promising results in capturing the violation possibility values in the order of $10^{-3}$. Furthermore, by employing the violation metrics for voltage magnitude and apparent branch flow, Tables \ref{risk_conf} and \ref{risk_bf_conf} present the 95\% confidence intervals derived from the MCS, alongside the average violation values yielded by our innovative methods. Notably, we observe that the mean violation degrees estimated through our proposed approach consistently fall within the 95\% confidence intervals established by the MCS technique.

\begin{table*}
\caption{The lower bound constraint violation probabilities ($10^{-2}$) of the voltage magnitude on the SouthCarolina-500 bus system}
\label{risk}
\centering
\scalebox{1}{
\begin{tabular}{|c|c|c|c|c|c|c|c|c|c|c|c|c|c|}\hline 
Bus index & MCS & LPF & SML &  KNN & RR & SVR &   FC & TPBNN & ResNet & Random & Data-driven & Linearized PF & Jacobian 
\\\hline
12 & 3.78 & 0.53 & 3.83 & 2.28 & 1.45 & 17.98 & 3.80 & 3.86 & 3.75 & 3.76 & \textbf{3.78} & 3.80 & \textbf{3.78}
\\\hline
13 & 3.85 & 0.65 & 4.01 & 2.38 & 1.56 & 18.48 & 3.88 & 3.88 & \textbf{3.86} & 3.88 & 3.88 & \textbf{3.86} & \textbf{3.86}
\\\hline
14 & 0.23 & 0 & 0.03 & 0 & 0 & 0 & 0.25 & 0.31 & \textbf{0.23} & \textbf{0.23} & \textbf{0.23} & \textbf{0.23} & \textbf{0.23}
\\\hline
15 & 0.95 & 0 & 0.61 & 0.13 & 0 & 2.70 & 0.97  & 1.15 & \textbf{0.95} & \textbf{0.95} & \textbf{0.95} & \textbf{0.95} & \textbf{0.95}
\\\hline
24 & 0.50 & 0 & 0.30 & 0 & 0 & 0.06 & 0.52 & \textbf{0.50} & 0.51 & 0.51 & 0.51 & 0.51 & 0.51
\\\hline
 
\end{tabular}
}
\end{table*}

\begin{table*}
\caption{The constraint violation probabilities ($10^{-2}$) of the apparent branch flow on the SouthCarolina-500 bus system.}

\label{risk_bf}
\centering
\scalebox{1}{
\begin{tabular}{|c|c|c|c|c|c|c|c|c|c|c|c|c|c|}\hline 
Branch index & MCS & LPF & SML &  KNN & RR & SVR &   FC & TPBNN & ResNet & Random & Data-driven & Linearized PF & Jacobian 
\\\hline
1 & 0.30 & 0.25 & 0.38 & 0 & 0.25 & 0.26 & 0.15 & 1.71 & 0.15 & 0.13 & 0.41 & \textbf{0.23} & 0.43
\\\hline
8 & 0.01 & \textbf{0.01} & \textbf{0.01} & 0 & \textbf{0.01} & 0 & \textbf{0.01} & \textbf{0.01} & \textbf{0.01} & \textbf{0.01} & \textbf{0.01} & \textbf{0.01} & \textbf{0.01}
\\\hline
9 &  0.01 & \textbf{0.01} & \textbf{0.01} & 0 & \textbf{0.01} & 0 & \textbf{0.01} & 0.05 & \textbf{0.01} & 0 & \textbf{0.01} & \textbf{0.01} & \textbf{0.01}
\\ \hline
18 & 0.16 & 0.15 &  0.55 & 0 & 0.15 & 0.13 & 0.13 & 0.11 & \textbf{0.16} & 0.13 & 0.15 & 0.15 & \textbf{0.16} 
\\\hline
19 & 0.63 & 0.38 & 1.13 & 0 & 0.48 & 0.28 & 0.53 & 0.53 & 0.41 & \textbf{0.63} & 0.60 & 0.67 & 0.53
\\\hline

\end{tabular}
}
\end{table*}

\begin{table*}
\caption{The 95\% confidence intervals of the voltage magnitude lower bound constraint violation values obtained by the MCS and the average violation values estimated by the proposed approaches on the SouthCarolina-500 bus system ($10^{-2}$).}
\label{risk_conf}
\centering
\scalebox{1}{
\begin{tabular}{|c|c|c|c|c|c|c|}\hline 

Bus index & MCS & Random & Data-driven & Linearized PF & Jacobian 
\\\hline
12 & [2.39, 3.08] & 2.75 & 2.75 & 2.75 & 2.74 
\\\hline
13 & [2.44, 3.12] & 2.77 & 2.77 & 2.80 & 2.78
\\\hline
14 & [0.62, 1.42] & 1.07 &  1.03 & 1.03 & 1.04 
\\\hline
15 & [1.86, 2.86] & 2.41 & 2.43 & 2.40 & 2.42
\\\hline
24 & [1.54, 2.48] & 1.99 & 1.97 & 1.96 & 1.99 
\\\hline

\end{tabular}
}
\end{table*}

\begin{table*}
\caption{The 95\% confidence intervals of the apparent branch flow upper bound constraint violation values obtained by the MCS and the average violation values estimated by the proposed approaches on the SouthCarolina-500 bus system.}
\label{risk_bf_conf}
\centering
\scalebox{1}{
\begin{tabular}{|c|c|c|c|c|c|c|}\hline 
Branch index & MCS & Random & Data-driven & Linearized PF & Jacobian
\\\hline
1 & [0.03, 0.09] & 0.05 & 0.07 & 0.08 & 0.07 
\\\hline
4 & [1.33, 1.39]  & 1.34 &  1.37 & 1.32 & 1.36 
\\\hline
10 & [6.07, 6.20]  & 6.11 &  6.19 & 6.15 &  6.15 
\\\hline
18 & [0.10, 0.37] &  0.22  &  0.21 & 0.19 & 0.19 
\\\hline
19 & [0.11, 0.26] &  0.20 & 0.22 & 0.18 & 0.19 
\\\hline

\end{tabular}
}
\end{table*}

In addition, one of the stopping criteria of the MCS method is the variance coefficient for the violation possibility to be smaller than 1\% \cite{Leite}. Within the simulations, accurate estimates of voltage magnitude violations and apparent branch flow violations necessitate a minimum of 2700 and 5100 samples, respectively. Thus, the cumulative computational time required to perform Monte Carlo simulations for power flow analysis across thousands of instances can be substantial. By contrast, our proposed methods can rapidly predict voltage phasors for thousands of data samples. 

\subsection{Computational Efficiency and Convergence Rate}
The simulations are implemented on an iMac with i7-8007 CPU and 32GB RAM and a Linux server with NVIDIA Tesla K20-5 GB GPU. The NN training is implemented via PyTorch 1.7.1 in Python 3.7. Table~\ref{time} shows the NN training time of each epoch. Residual building blocks need a slightly longer time due to the extra propagation of the shortcut connection layer. Table~\ref{time_test} shows our proposed approaches significantly reduce the total computational time compared with the MCS and QMC (with 3000 sampling points) algorithms. 

\begin{table}
\caption{Computation training time comparison of each epoch for different cases (seconds).}
\label{time}
\centering
\begin{tabular}{c|c|c|c|c}\hline 

Cases  & FC & TPBNN & ResNet & Proposed approaches \\\hline

\multirow{1}{*}{30} & 0.87 & 2.32 & 1.13 & 1.07 \\ \hline

\multirow{1}{*}{118} & 1.55 & 4.11 & 1.90  & 1.91  \\  \hline

\multirow{1}{*}{300} & 2.13 & 6.27 & 1.97  & 2.37 \\  \hline

\multirow{1}{*}{500} & 3.57 & 15.66  & 4.57 & 3.74 \\  \hline

\multirow{1}{*}{1354} & 4.58 & -  & 6.10 & 5.77 \\  \hline

\end{tabular}
\end{table}

\begin{table*}
\caption{Computation testing time comparison for different cases (seconds) and the calculation time ratio of MCS to our proposed approaches. }
\label{time_test}
\centering
\begin{tabular}{c|c|c|c|c|c|c|c|c|c|c|c|c}\hline 

Cases & MCS & QMC & LPF & SML & KNN & RR & SVR & FC & TPBNN & ResNet & Proposed approaches & Acceleration ratio \\\hline

\multirow{1}{*}{30} & 7.88 & 6.34 & 0.01 & 0.09 & 0.08 & 0.00 & 71.28 & 0.00 & 0.01 & 0.00 & 0.00 & 1136  \\ \hline

\multirow{1}{*}{118} & 14.95 & 9.67 & 0.04 & 0.44 & 0.36 & 0.02 & 173.83 & 0.00 & 0.22 & 0.00 & 0.00 & 1657 \\  \hline

\multirow{1}{*}{300} & 32.11 & 21.96 & 0.22 & 0.63 & 0.96 & 0.05 & 3028.93 & 0.01 & 0.62 & 0.01 & 0.01 & 3642 \\  \hline

\multirow{1}{*}{500} & 48.48 & 26.60 & 0.34 & 0.93 & 2.77 & 0.11 & 322.13 & 0.01 & 1.87 & 0.01 & 0.01 & 2921 \\  \hline

\multirow{1}{*}{1354} & 125.47 & 51.60 & 2.75 & 6.59 & 9.34 & 0.75 & 127.81 & 0.04 & - & 0.05 & 0.04 & 2598 \\  \hline

\end{tabular}
\end{table*}

The learning rate affects the loss convergence rate of the NN training process. A significant learning rate helps fast convergence but may lead to the NN weights converging to a suboptimal solution. Hence, we adopt a relatively small learning rate of $10^{-4}$. Note that this small learning rate is only used in this part to indicate the convergence properties. The MSE evolution of the training loss in the first 500 epochs is shown in Fig.~\ref{fig:mse_loss30} and Fig.~\ref{fig:mse_loss118}. After training the first epoch, the MSEs of our proposed approaches are more than two orders less than that of others. Besides, even after training 500 epochs, other NN-based methods cannot achieve the same loss level as ours. Therefore, our proposed approaches have significant advantages over others regarding the convergence rate, which can be attractive in the face of a limited training time.

In addition, three designed initialization schemes converge faster than the random initialization. Therefore, we show how designed initialization methods influence the NN weights update during the training process. Fig.~\ref{fig:imgshow_initial_random} shows the randomly distributed parameters of the shortcut connection layer. After training 500 epochs, the pattern of parameters is still very random, as shown in Fig.~\ref{fig:imgshow_later_random}. In contrast, Fig.~\ref{fig:imgshow_all} shows that the pattern of the updated parameters after training is still quite similar to that of the initial parameters. This phenomenon indicates that these learnable parameters are finely updated based on the initial weights during the training process. Therefore, we conclude that our well-designed initial weights play a critical role in NN training.

\section{Conclusion}\label{sec:conclusion}
This paper proposes a novel residual learning NN framework with three different initialization schemes to conduct rapid PF analysis, which can significantly reduce the total computational time in PPF analysis. Traditional PF analysis relies on the NR solver to solve the AC-PF equations iteratively until convergence. The widely installed PMUs and SCADA systems can collect abundant measurements, which motivates the necessity of considering NNs for real-time PF analysis. 

Inspired by the residual building block, we introduce a shortcut connection linear layer between the input power injections and output voltage phasors to the MLP structure. Our proposed framework aims to learn the non-linear correction to the linearized AC-PF equations instead of directly dealing with the original non-linear AC-PF equations. In addition, the absolute values of voltage angle differences between connected buses are typically small while the voltage magnitudes are slightly perturbed around 1 per unit. Based on this property, we develop three initialization schemes for different scenarios. Two model-based schemes (linearized PF and Jacobian methods) require knowledge of network topology and line parameters. If this information is missing or inaccurate, the data-driven approach will be a good choice. Extensive simulation results show that our proposed approaches improve the estimation accuracy and significantly speed up the training when compared with the existing methods. 

\nocite{*}

\bibliographystyle{IEEEtran}
\bibliography{ref_SJ,IEEEabrv}

\end{document}